\documentclass[pdflatex,sn-basic]{sn-jnl}
\AtBeginDocument{%
  }

\usepackage{amsmath}
\usepackage{listings}
\usepackage[T1]{fontenc}
\usepackage{textcomp}
\usepackage[most]{tcolorbox}
\usepackage{graphicx}
\usepackage{booktabs}
\usepackage{tabularx}
\usepackage{array}
\usepackage{svg}
\usepackage{algorithm}
\usepackage{algpseudocode}
\usepackage{amsmath}

\usepackage{booktabs}
\usepackage{tabularx}
\usepackage{makecell}
\usepackage{threeparttable}
\usepackage{xcolor}
\usepackage{framed}
\usepackage{tikz}
\usepackage{listings}
\usepackage{graphicx}
\usepackage{enumitem}
\usepackage{tcolorbox}
\usepackage{subcaption}
\usepackage{adjustbox}
\usepackage{booktabs}
\usepackage{multirow}
\tcbuselibrary{listings,skins,breakable}

\usepackage{colortbl}
\usepackage{pifont}
\usepackage{ragged2e}

\usepackage[normalem]{ulem}

\newcolumntype{P}[1]{>{\RaggedRight\arraybackslash}p{#1}}

\definecolor{codebg}{RGB}{250,250,250}
\definecolor{codeframe}{RGB}{210,210,210}
\definecolor{titlebg}{RGB}{238,238,238}
\definecolor{commentgray}{RGB}{110,110,110}
\definecolor{violhl}{RGB}{255,239,222}

\definecolor{ericrewritegreen}{RGB}{235,250,235}

\definecolor{llmtokengray}{gray}{0.95}

\newcommand{\cmark}{\ding{51}}

\newcommand{\yesbadge}[1]{\ding{51}}
\newcommand{\nobadge}[1]{\ding{55}}

\newlength{\rotgrouprowheight}

\lstdefinestyle{cviol}{
  language=C,
  basicstyle=\ttfamily\footnotesize,
  numbers=none,
  frame=none,
  showstringspaces=false,
  columns=fullflexible,
  keepspaces=true,
  tabsize=2,
  breaklines=true,
  breakatwhitespace=true,
  commentstyle=\color{commentgray},
  escapeinside={(*@}{@*)}
}

\newcommand{\viol}[1]{%
  \begingroup\setlength{\fboxsep}{0pt}%
  \colorbox{violhl}{\strut\hspace{2pt}#1\hspace{2pt}}%
  \endgroup
}

\newcommand{\tech}{\textsc{KLEECopilot}}
\newcommand*\circled[2]{%
  \definecolor{currentcolor}{HTML}{#1}
  \tikz[baseline=(C.base)]
    \node[
      draw, 
      circle, 
      fill=currentcolor,
      text=white, 
      inner sep=1pt,
      font=\small
    ](C) {#2};\!
}

\newtcolorbox{summarybox}{
  enhanced,
  colback=white,
  colframe=black,
  boxrule=0.8pt,
  boxsep=0pt,
  left=4pt,
  right=4pt,
  top=4pt,
  bottom=4pt,
  before skip=6pt,
  after skip=6pt,
  before upper={\textbf{Conclusion:}\enskip}
}

\usepackage{mdframed}

\newmdenv[
    topline=true,
    bottomline=true,
    rightline=false,
    leftline=false,
    linewidth=1.5pt,
    skipabove=8pt,
    skipbelow=8pt,
    innerleftmargin=0pt,
    innerrightmargin=0pt,
    frametitle={Conclusion:},
    frametitlefont=\bfseries
]{conclusion}

\definecolor{llvmbg}{RGB}{248,248,248}
\definecolor{llvmkw}{RGB}{30,74,150}     
\definecolor{llvmty}{RGB}{0,110,110}     
\definecolor{llvmfl}{RGB}{120,60,140}    
\definecolor{llvmcm}{RGB}{110,110,110}   
\definecolor{llvmlit}{RGB}{120,70,20}    

\lstdefinelanguage{LLVM}{
  morecomment=[l]{;},
  sensitive=true,
  morekeywords={
    add,sub,mul,udiv,sdiv,urem,srem,shl,lshr,ashr,and,or,xor,
    load,store,alloca,phi,call,ret,br,switch,select,
    getelementptr,icmp,fcmp,bitcast,ptrtoint,inttoptr,trunc,zext,sext
  },
  morekeywords=[2]{void,i1,i8,i16,i32,i64,i128,float,double,ptr},
  morekeywords=[3]{nsw,nuw,inbounds,align,exact,fast,ninf,nnan,nsz,arcp,contract,reassoc},
}

\definecolor{taodonecolor}{RGB}{150,80,150}
\definecolor{ericdonecolor}{RGB}{180,80,80}

\AtBeginDocument{%
  }

\newtcblisting{promptbox}{
  listing only,
  breakable,
  colback=white,
  colframe=black,
  boxrule=0.4pt,
  enhanced,
  listing options={
    basicstyle=\ttfamily\footnotesize,
    breaklines=true,
    columns=fullflexible,
    numbers=none,
    frame=none,
    keywordstyle=\ttfamily,
    identifierstyle=\ttfamily,
    commentstyle=\ttfamily,
    stringstyle=\ttfamily,
    showstringspaces=false
  }
}

\newtcolorbox{chatwindow}{
  enhanced,
  colback=white,
  colframe=black!20,
  boxrule=0.5pt,
  arc=4pt,
  width=0.9\linewidth,
  left=1pt,right=1pt,top=1pt,bottom=1pt,
  fontupper=\fontsize{8pt}{1.6pt}\selectfont,
}

\newtcolorbox{humanmsg}{
  enhanced,
  breakable,
  colback=white,
  colframe=black!15,
  boxrule=0.4pt,
  arc=2pt,
  left=4pt,right=4pt,top=3pt,bottom=3pt
}

\newtcolorbox{botmsg}{
  enhanced,
  breakable,
  colback=black!3,
  colframe=black!15,
  boxrule=0.4pt,
  arc=2pt,
  left=4pt,right=4pt,top=3pt,bottom=3pt
}

\begin{document}

\title{Directed Symbolic Execution for Vulnerability Discovery: An LLM-Guided Approach in KLEE}





\author[1]{\fnm{Lingfeng} \sur{Chen}}
\email{lingfeng@ait.kyushu-u.ac.jp}

\author*[1]{\fnm{Tao} \sur{Xiao}}
\email{xiao@ait.kyushu-u.ac.jp}

\author[1]{\fnm{Masanari} \sur{Kondo}}
\email{kondo@ait.kyushu-u.ac.jp}

\author[1]{\fnm{Yasutaka} \sur{Kamei}}
\email{kamei@ait.kyushu-u.ac.jp}

\affil*[1]{\orgname{Kyushu University},
\orgaddress{\city{Fukuoka}, \country{Japan}}}


\abstract{

Symbolic execution effectively discovers security violations but suffers from path explosion. Engines like \textsc{KLEE} therefore use path prioritization heuristics to order state exploration, typically optimizing code coverage. However, path prioritization can become trapped in cyclic control-flow regions, where repeated branching consumes the exploration budget before exploration reaches vulnerable code beyond these cyclic regions. We propose \tech{}, a Large Language Model (LLM)-guided directed symbolic execution approach built on \textsc{KLEE}. \tech{} uses LLMs to mark potentially vulnerable code and guide path prioritization. It also integrates loop-exit prioritization to escape potentially non-vulnerable cycles and progress toward deeper vulnerabilities. Compared with baselines such as \textsc{Empc}, \tech{} improves basic block coverage by 42.24\% and line coverage by 125.82\%. It discovers 1,335 total violations and 87 unique violations, outperforming the second-best baseline by 32.2\% in total violations and \textsc{Empc} by 24.3\% in unique violations. Although \tech{} is sensitive to model family, it exhibits only marginal sensitivity to model scale, supporting the efficacy of integrating security semantics and loop-exit prioritization. Ablation studies further show that individual components contribute to effectiveness: alternative configurations involving searchers, internal components, marking sources, and prompt variants yield only 54--61 unique violations, while \tech{} maintains competitive code coverage.
}


\maketitle

\section{Introduction}
Symbolic execution~\citep{king1976symbolic} is a powerful program analysis technique that finds security violations by exploring program paths with symbolic inputs---algebraic variables that represent all possible values, instead of concrete values, to find potential vulnerabilities. Engines such as \textsc{KLEE}~\citep{cadar2008klee} made symbolic execution practical for real-world C/C++ programs by executing LLVM bitcode, 
and by generating concrete tests from path constraints. This makes \textsc{KLEE} particularly suitable for automated bug and security violation discovery: (i) it analyzes programs without source-code modifications, while automatically generating concrete test inputs that make discovered failures or security violations easy for testers to validate and reproduce; (ii) unlike concrete testing that follows a single execution path per run, \textsc{KLEE} can simultaneously explore multiple execution paths; and (iii) because \textsc{KLEE} uses symbolic variables, it does not need to enumerate all possible concrete values for each path, allowing it to explore more program behaviors within a limited time budget. As a result, \textsc{KLEE} has remained a widely used symbolic execution engine in both
academia and industry over the past decade~\citep{cadarstudy, chipounov2011s2e,klee2019engine,zhang2022random_or_heuristic_klee,he2021learch,sun2024cgs,busse2024ssle,empc2025}.

However, scaling symbolic execution to complex programs remains difficult. Whenever the program reaches a branch point dependent on symbolic inputs, such as a conditional branch or a loop branch, symbolic execution forks into two distinct paths corresponding to the two feasible branch directions. This causes the total number of paths to grow exponentially with program complexity. Consequently, \textbf{path explosion stands as the primary scalability challenge in symbolic execution and in \textsc{KLEE}}~\citep{cadar2013symbolic,baldoni2018survey,bohme2025software,bugrara2013rsd,avgerinos2014veritesting,trabish2018chopped,he2021learch,sun2024cgs}.

Prior studies have primarily focused on \textbf{path prioritization} as a way to mitigate path explosion during symbolic execution~\citep{li2013steering,he2021learch,zhang2022random_or_heuristic_klee,sun2024cgs,empc2025}. It selects one path from many pending paths to continue execution, typically aiming to increase code coverage. While higher coverage may increase the chance of reaching potentially vulnerable locations, the searcher (\textsc{KLEE}'s strategy for selecting the next execution path) does not know where such locations are before exploration. A more targeted form of path prioritization is directed symbolic execution, which addresses this limitation by prioritizing paths leading toward specified locations, thereby reducing unnecessary exploration when the targets are already known~\citep{ma2011directed,yang2014directed,babic2011sandwich,gerasimov2018directed,busse2022combining,tu2024vital}. However,  prior directed symbolic execution is mainly designed to validate given targets, rather than to broadly explore the entire program for security violations. Another gap is how to avoid spending the budget in cyclic control-flow regions, such as infinite loop~\citep{klee777issue,klee831issue}, where repeated branching may consume a large amount of exploration budget before exploration reaches vulnerable paths beyond these cyclic regions~\citep{xiao2013loopstudy,gao2018loop,busse2024ssle}. These issues motivate our design from the following two complementary perspectives.

\textbf{Lack of Vulnerability-Oriented Code Semantics.}
The above-mentioned studies have used different target sources, including changed code regions~\citep{yang2014directed}, static analysis warnings~\citep{babic2011sandwich,gerasimov2018directed}, and static analysis error traces~\citep{busse2022combining}. Their targets are usually derived from specific program locations, predefined rules, or constrained by the capabilities of static analyzers. Precise analysis may miss cases that are difficult to capture statically, while broad rule-based matching may introduce excessive warning noise~\citep{bessey2010few,johnson2013why,sadowski2018lessons}. Therefore, they are more suitable for validating given targets than for broadly discovering security violations across the analyzed program. Moreover, recent work such as \textsc{VITAL} extends directed symbolic execution by using pointer analysis~\citep{tu2024vital}, but it is still limited to unsafe-pointer-related security violations. We tackle this problem from a semantics-based perspective: instead of relying on semantically shallow predefined rules or constrained analysis algorithms, we leverage the LLM's code understanding and security knowledge to mark potentially vulnerable code locations across broader vulnerability types.

\textbf{Difficulty of Loop Handling.} 
Path prioritization becomes especially difficult in cyclic regions, where repeated branching may consume a large amount of exploration budget before deeper vulnerable paths are reached. \textsc{KLEE} does not provide built-in optimization specifically for this difficulty. Therefore, prior work has explored loop-specific techniques. \citet{gao2018loop} reduce the number of states generated by \textsc{KLEE} inside loops through loop path reduction by \textsc{KLEE} state pruning, while \citet{kapus2019loops} propose loop summarization for specific classes of string loops. Our motivation is different. We do not simply force symbolic execution to leave every loop. Instead, we treat loop handling as a violation-aware prioritization problem: if a cyclic region contains potentially vulnerable
code locations, we allow exploration to continue inside the region toward those locations. Otherwise, we prioritize paths that leave the cyclic region to avoid redundant loop-heavy exploration. This design choice aligns with our objective of maximizing security violation discovery under limited exploration resources.

To address these two limitations, we propose \tech{}, an LLM-guided directed symbolic execution approach built on top of \textsc{KLEE}. First, it uses a Large Language Model (LLM), which has shown promise in capturing semantics related to security violations and improving understanding of the context and rationale behind vulnerabilities~\citep{du2024vulllm}, to produce markings for potentially vulnerable code regions. These markings provide semantic guidance related to security violations for path prioritization and help avoid blind exploration. Second, based on these markings, \tech{} further incorporates loop-exit path prioritization by collapsing each cyclic region into a virtual node only for cyclic-region classification. This allows \textsc{KLEE} to distinguish between cyclic regions that contain potentially vulnerable markings and those that do not, continuing exploration inside marked cyclic regions while prioritizing exits from unmarked ones.

Our main \textbf{contributions} are as follows:
\begin{itemize}
  \item We present \tech{}, an LLM-guided vulnerability-oriented directed symbolic execution approach built on top of \textsc{KLEE}. It uses LLM-generated, source-level security violation markings as target guidance and combines them with violation-aware loop-exit prioritization to steer symbolic execution toward security-critical code regions.
  \item We conduct a comprehensive evaluation of \tech{} against 13 baseline methods and seven experimented LLMs, as well as 12 ablation variants covering four configurations. Our results demonstrate that \tech{} achieves the best unique-violation discovery while maintaining competitive, and sometimes best, code coverage.
  \item We release \tech{} in the replication package~\citep{repulication_package} and a benchmark suite together with a fully containerized experimental environment to support reproducibility.  
\end{itemize}

\section{Background}

\begin{figure}[!hbp]
\centering
\includegraphics[width=0.80\linewidth]{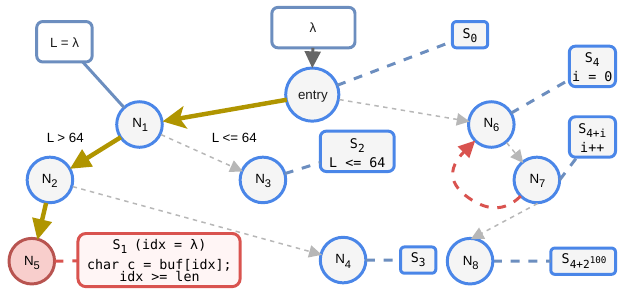}
\caption{Path explosion. Repeated loop branching increases
the number of symbolic states, while reaching vulnerable code
may require selecting a loop-body or loop-exit path}
\label{fig:overview-path-explosion}
\end{figure}

\subsection{Path Explosion Problem}
\textsc{KLEE} works as an instruction-level symbolic interpreter over LLVM bitcode of the program. During execution, it maintains a set of active execution states, where each state represents one path under exploration together with its current instruction location, stack, address space, and constraints on symbolic variables accumulated along the path. When execution reaches a branch whose condition depends on symbolic inputs, each feasible branch direction creates a corresponding successor state. As exploration proceeds, repeated branching continuously enlarges the set of candidate paths, leading to the path explosion problem.

Loops make this problem more severe. On the right side of Figure~\ref{fig:overview-path-explosion}, the loop branch at $N_7$ may be revisited across iterations. In \textsc{KLEE}, each additional iteration corresponds to a new execution path rather than the same path repeated, because the path history is extended by another loop traversal. In particular, after each iteration, loop-carried variables and branch conditions keep adding new constraints to the current state. As a result, even when execution revisits the same loop branch node, the associated path constraints differ from those of earlier iterations. Thus, what appears to be the same loop path at the control-flow level may correspond to different states in \textsc{KLEE}, which causes loops to generate many active states (e.g., $S_{4+2^{100}}$ in Figure~\ref{fig:overview-path-explosion}) and quickly enlarge the search space.

More generally, this issue is not confined to regular loops, but also manifests in other cyclic control-flow structures, including irregular cycles within a function and recursive cycles across functions. Consequently, cyclic control-flow structures can generate a large number of simultaneously active states, thereby more rapidly expanding the associated search space.

\subsection{Reaching Vulnerable Paths}

The left side of Figure~\ref{fig:overview-path-explosion} illustrates the main difficulty of symbolic execution from the perspective of reaching vulnerable paths. Starting from the symbolic input $\lambda$, \textsc{KLEE} begins at the entry node and forks execution whenever a branch is feasible under the current path constraints. Before the vulnerable access at \(N_5\) is reached, the execution has already forked into several candidate paths, i.e., \(\textit{entry}\!\rightarrow\!N_1\!\rightarrow\!N_2\!\rightarrow\!N_5\), \(\textit{entry}\!\rightarrow\!N_1\!\rightarrow\!N_2\!\rightarrow\!N_4\), and \(\textit{entry}\!\rightarrow\!N_1\!\rightarrow\!N_3\). However, only a small subset of these paths can actually continue toward vulnerable access (\(N_5\)), while the others consume a time budget without helping to reach it. As branching continues, this effect quickly accumulates and makes vulnerable paths increasingly difficult to reach with a limited time budget.

This difficulty further highlights how crucial path prioritization is for the execution of \textsc{KLEE}. Although \textsc{KLEE} can confirm a violation once a vulnerable path is reached, it does not know which following branches are more vulnerable to reaching such paths. Our goal is therefore not to change how violations are validated, but to guide exploration toward paths that are more likely to lead to vulnerable code. Once such a path is reached, \textsc{KLEE} can check whether the current path constraints satisfy the corresponding violation condition and send a \path{ktest} file for replay validation. We also use replay as validation to examine the effectiveness of \tech{} in terms of discovering security violations in RQ2.

\section{Related Work}
\label{sec:rw}
Recent work relevant to our study can be organized into the following three research directions.

\textbf{Path Prioritization in \textsc{KLEE}.}
This strategy in \textsc{KLEE} has been studied using execution history, machine learning, concrete guidance, and graph-based structural approaches. Representative examples include \textsc{SGS}~\citep{li2013steering}, \textsc{Learch}~\citep{he2021learch}, \textsc{CGS}~\citep{sun2024cgs}, and \textsc{Empc}~\citep{empc2025}; we will further discuss these approaches in Section~\ref{sec:setup}. Unlike these approaches, \tech{} introduces source-level semantic security violation guidance based on LLM-generated marking into \textsc{KLEE}'s path prioritization. It also considers cyclic regions, such as a loop that can cause path explosion during its iteration~\citep{xiao2013loopstudy}.

\begin{center}
\captionof{table}{Synthesis of related work in directed symbolic execution.}
\label{tab:related-works}

\begin{threeparttable}
\begin{tabularx}{\textwidth}{
  lc>{\RaggedRight\arraybackslash}X
    >{\RaggedRight\arraybackslash}X}
\toprule[0.8pt]
\textbf{Study} &
\textbf{KLEE} &
\textbf{Marking Source} &
\textbf{Scenario} \\
\midrule

\textbf{Our approach} &
\cmark &
LLM-generated markings &
ASan/UBSan violation discovery \\

\specialrule{0.25pt}{2pt}{2pt}

\citet{ma2011directed} &
 &
-- &
specific program location discovery \\

\citet{babic2011sandwich} &
 &
static analyzer &
warning validation \\

\citet{yang2014directed} &
 &
code changes &
iterative security testing \\

\citet{yao2017statsym} &
\cmark &
runtime logs &
violation discovery \\

\citet{gerasimov2018directed} &
 &
static analyzer &
warning validation \\

\citet{busse2022combining} &
\cmark &
static analyzer &
warning validation \\

\citet{tu2024vital} &
\cmark &
pointer analysis &
unsafe-pointer violation discovery \\

\bottomrule[0.8pt]
\end{tabularx}

\begin{tablenotes}
\footnotesize
\item KLEE = whether the approach is implemented on top of \textsc{KLEE}.
\end{tablenotes}
\end{threeparttable}
\end{center}

\textbf{Directed Symbolic Execution.}
As shown in Table~\ref{tab:related-works}, directed symbolic execution studies how to steer symbolic execution toward specified program locations, static analysis warning locations, or vulnerability-related code locations. \citet{ma2011directed} introduced directed symbolic execution strategies such as Shortest Distance Symbolic Execution (SDSE), which prioritizes paths by their interprocedural distance in the control-flow graph to a target. Directed incremental symbolic execution applies directed exploration to change-impact analysis by focusing exploration only on paths affected by code changes for iterative testing~\citep{yang2014directed}.

Vulnerability-oriented directed symbolic execution has also used static analysis to assist existing testing methods, including symbolic execution. Prior work uses custom static analysis to generate vulnerability warnings or paths that could trigger potential vulnerabilities, and then uses symbolic execution as a tool to validate them. \citet{babic2011sandwich} uses symbolic execution to validate static analysis warnings. \citet{busse2022combining} further uses symbolic execution to validate paths that could potentially trigger vulnerabilities reported by Infer~\citep{infer}, a widely adopted industry-class static analyzer. \citet{yao2017statsym} uses symbolic execution to validate logs from runtime execution. These studies are similar to \tech{}, but they apply symbolic execution as a validation tool for one target at a time, instead of using it to prioritize paths across multiple targets. The work most closely related to \tech{} is \textsc{VITAL}, which extends directed symbolic execution with pointer analysis; however, its target scenario is limited to unsafe-pointer-related security violations~\citep{tu2024vital}.

Furthermore, static analyzers used in directed symbolic execution usually rely either on constrained analysis algorithms, such as data-flow analysis, or on manually specified rules and patterns. The resulting warnings or marked locations are constrained by the underlying analysis models, rules, and vulnerability types. Apart from prior work, \tech{} uses LLMs to directly mark potentially vulnerable source-code locations based on the ASan and UBSan security definitions in our prompt and the LLM's security knowledge. Moreover, we do not use symbolic execution as a validation tool. Instead, we prioritize paths in \textsc{KLEE} toward multiple LLM-generated markings, and then use replay to confirm security violations.

\textbf{LLMs for Vulnerability Localization and LLM-Assisted KLEE.}
Recent work has applied LLMs to vulnerability detection and localization, often with a richer semantic or structural context~\citep{du2024vulllm,wu2024vffinder,lekssays2025llmxcpg,tian2025envul}. Moreover, previous evaluations show that LLM-based vulnerability analysis remains sensitive to model family, parameter scale, and input or inference conditions~\citep{ullah2024llms,lin2025comparative,sun2024llm4vuln}. Unlike these studies, \tech{} does not use LLMs for vulnerability detection, but instead uses potentially vulnerable code regions produced by LLMs as a semantic guide for symbolic exploration.

Existing work combining LLMs with \textsc{KLEE} remains limited. \citet{bouras2026gordian} uses LLM-generated ghost code to help \textsc{KLEE} handle code fragments that are difficult for the solver, rather than to guide path prioritization. \citet{xu2024symbolic} and \citet{wang2025nexusym} use LLM-generated test cases to assist symbolic execution, rather than guiding path prioritization within \textsc{KLEE}. \citet{ma2026symtee} use an LLM to complete code slices into programs that \textsc{KLEE} can execute, rather than to guide its exploration. Therefore, none of them can be compared in our setting.

\section{\tech{}}
\label{sec:method}

\begin{figure}[t]
\centering
\includegraphics[
  trim=12bp 12bp 12bp 12bp,
  clip,
  width=0.80\linewidth
]{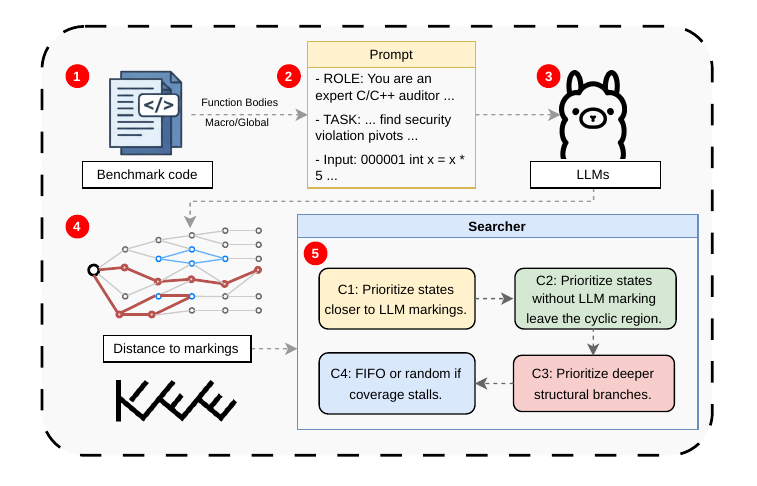}
\caption{\tech{} architecture.
LLM-generated markings guide distance computation
and state prioritization in \textsc{KLEE}}
\label{fig:overview-architecture}
\end{figure}

In this section, we present \tech{}, a Large Language Model (LLM)-guided vulnerability-oriented directed symbolic
execution approach. 
Specifically, it consists of three phases shown in Figure~\ref{fig:overview-architecture}: 
(i) identifying potential vulnerable locations at the line-level granularity (\circled{FF0000}{1}--\circled{FF0000}{3}); (ii) incorporating the LLM-generated markings into the structural analysis, providing loop-exit and path-level guidance for effective prioritization (\circled{FF0000}{4}); and (iii) adopting necessary modifications to \textsc{KLEE} and introducing a custom searcher for prioritization (\circled{FF0000}{5}). 

\subsection{LLM-Assisted Potential Vulnerability Localization}
\label{sec:phase1}

\begin{figure}[t]
\centering
\includegraphics[width=\columnwidth]{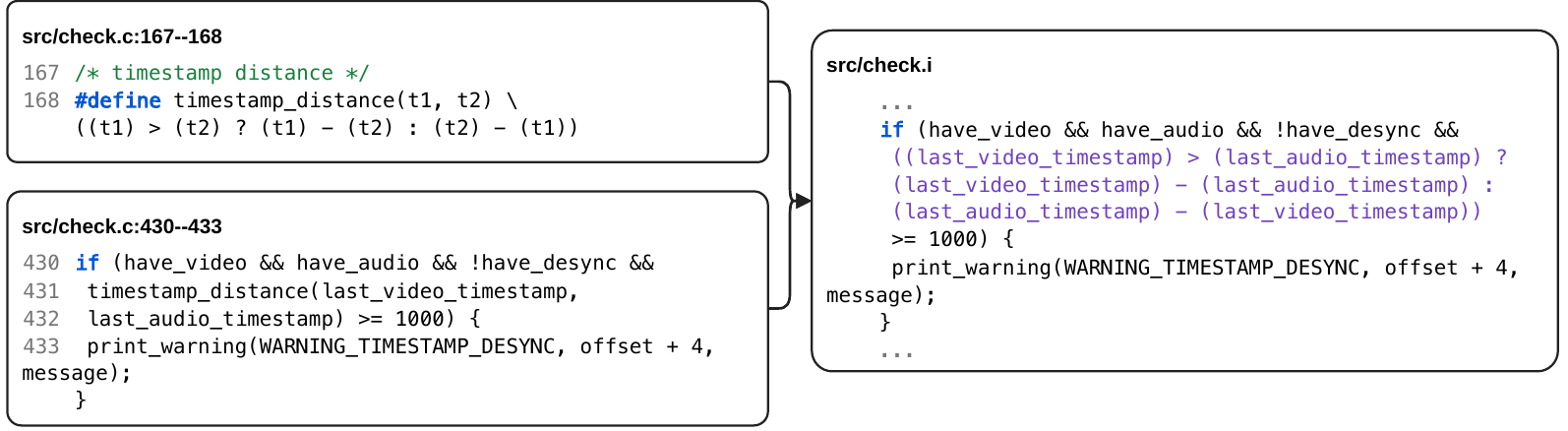}
\caption{After pre-processing, the source code is macro-free, which is consistent with the LLVM bitcode binary and our \textsc{KLEE} testing environment}
\label{fig:preprocessed-macro-expansion}
\end{figure}

We use an LLM to identify \emph{potentially vulnerable locations} at line-level granularity within a given program. Our prompt is structured into a fixed system prompt (such as marking strategies) and a user prompt (such as the benchmark program’s code slice). To keep the LLM-generated markings of source code consistent with the final program, we first use the C/C++ compiler itself to pre-process the source files using the same compilation environment used to generate the program, as shown at \circled{FF0000}{1} in Figure~\ref{fig:overview-architecture}. This step expands macros (such as \textit{timestamp\_distance} shown in Figure~\ref{fig:preprocessed-macro-expansion}) and resolves conditional compilation before LLM-generated marking, so the LLM can scan the code that is actually compiled and explored by \textsc{KLEE}. Figure~\ref{fig:preprocessed-macro-expansion} illustrates this process with a macro expansion process from a benchmark program \textit{flvmeta} in later evaluation: a macro invocation in the source file is transformed into the corresponding code in the preprocessed file.

We identify the source code files to be analyzed using the existing debug metadata in the LLVM bitcode. Specifically, we collect the source code paths referenced by the metadata and exclude all other files. We then parse the pre-processed source files using Tree-sitter and extract function bodies as input. Following prior work~\citep{ahmed2025secvuleval,chen2024llm4fpm}, we enhance accuracy by providing more than just the function body to the model. Specifically, we also include the relevant context, such as global definitions associated with the target function. Furthermore, to mitigate hallucinations, we annotate each line in the function body with its source line number before sending it to the LLM~\citep{hossain2024deepdive} (\circled{FF0000}{2}). Then, both system and user prompts are sent to the LLM (\circled{FF0000}{3}).

\begin{figure}[t]
  \centering
  \includegraphics[width=\columnwidth]{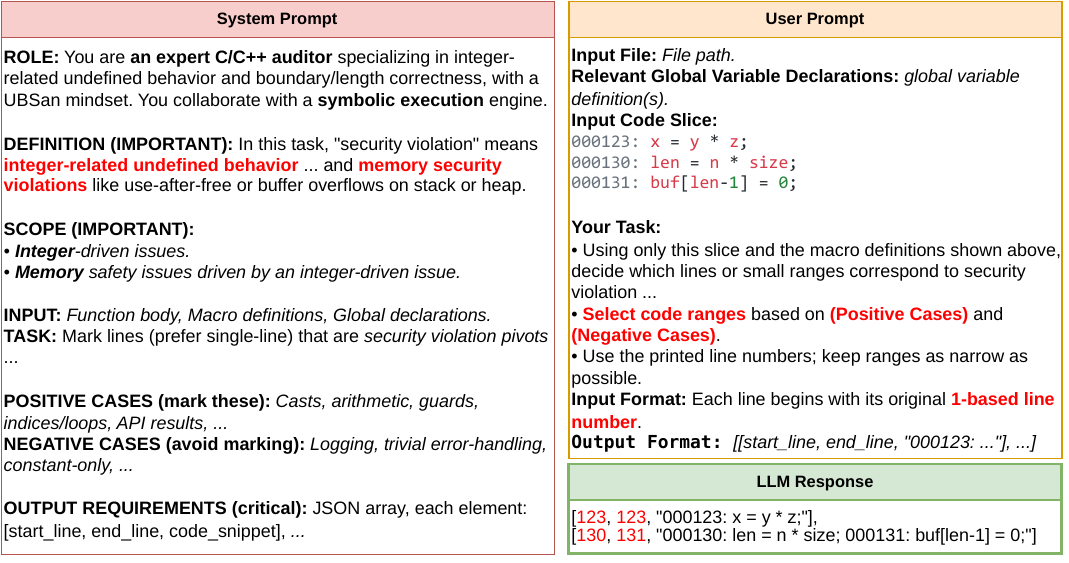}
  \caption{Prompt template and response for identifying potentially vulnerable locations within a given program}
  \label{fig:prompt}
\end{figure}

Figure~\ref{fig:prompt} shows a simplified prompt template used to identify potentially vulnerable locations. Following established prompt engineering practices~\citep{openai,liu2023pre}, we construct the system prompt to include the role of the LLM, task and term definitions, the scope of violations, marking guidelines, and the specific format of input and output. In particular, we encourage the LLM to mark positive cases where input-controlled values influence memory safety-related code or potential undefined behaviors. Memory safety concerns the integrity of memory access (e.g., ensuring pointers remain within allocated boundaries), while undefined behaviors refer to cases whose results are not specified by the language standard, such as integer overflows or null pointer dereferences. Conversely, we instruct them to avoid marking negative cases: (i) pure logging since it does not contribute vulnerability-relevant behavior~\citep{peng2024angel}, (ii) straightforward error-handling since it only returns, aborts, or propagates an error without introducing a new security-relevant pivot~\citep{wu2021hero}, and (iii) input-independent constant computations since they do not participate in vulnerability-relevant behavior~\citep{peng2024angel}. Positive cases defined in the system prompt are based on a predefined list of security violations from the following two sources.

\begin{itemize}
\item Inspired by previous studies~\citep{Ni2026ExtensiveStudy,Foulefack2026DomainAware,Ni2026MVulDPlus}, we employ top-10 CWEs in the real-world C/C++ vulnerability dataset, MegaVul~\citep{megavul}, to ground our violation types in the prompt: \emph{Integers}~\citep{cwe738} (CWE-190), \emph{Memory Safety}~\citep{cwe1399} (CWE-787, CWE-119, CWE-416, CWE-125), \emph{Memory Management}~\citep{cwe742} (CWE-476), \emph{Improper Input Validation}~\citep{cwe1406} (CWE-20), \emph{Sensitive Information Exposure}~\citep{cwe1417} (CWE-200), \emph{Missing Release of Memory after Effective Lifetime}~\citep{cwe401} (CWE-401), and \emph{Race Condition}~\citep{cwe362} (CWE-362).

\item In our study, \textsc{KLEE} is used to generate concrete test inputs, and the resulting violations are validated during replay on sanitizer-instrumented binaries. Following previous studies~\citep{empc2025,he2021learch}, we use two major sanitizer families: AddressSanitizer (ASan)~\citep{asandoc} and UndefinedBehaviorSanitizer (UBSan)~\citep{ubsandoc}. ASan detects out-of-bounds accesses and use-after-free, while UBSan detects undefined behaviors such as null pointer dereference and integer overflow. Therefore, we include: \emph{Out-of-bounds Write}~\citep{cwe1399} (CWE-787), \emph{Out-of-bounds Read}~\citep{cwe1399} (CWE-125), \emph{Improper Restriction of Operations within the Bounds of a Memory Buffer}~\citep{cwe1399} (CWE-119), and \emph{Use After Free}~\citep{cwe1399} (CWE-416) for ASan-enabled replay, together with \emph{Integer Overflow or Wraparound}~\citep{cwe738} (CWE-190) and \emph{NULL Pointer Dereference}~\citep{cwe742} (CWE-476) for UBSan-validated cases.
\end{itemize}

Finally, before symbolic execution, we collect a JSON array of LLM-generated and line-numbered markings. 

\subsection{Loop-Exit and LLM-Guided Path Prioritization}
\label{sec:phase2}

In \tech{}, \textsc{KLEE} is integrated with a combination of loop-exit and LLM-guided path prioritization to navigate the exploration. To support this, we first build an inter-procedural control-flow graph (ICFG) ~\citep{reps1995interprocedural}. We then compute Strongly Connected Components (SCC) using Tarjan's algorithm~\citep{tarjan1972} to identify cyclic regions, including standard loops (\path{for} or \path{while}), and inter-procedural cycles across functions (e.g., recursive calls). The SCC collapse is not used to replace the original ICFG for distance computation. Instead, every ICFG edge is assigned \textbf{one} unit weight, and the distance between two nodes is the shortest number of ICFG edges between them. Thus, loop collapse does not change the distance between ordinary ICFG nodes. Its purpose is to classify cyclic regions according to whether they contain any LLM-generated marking: \tech{} continues exploration inside marked cyclic regions, while prioritizing exits from unmarked cyclic regions. We compute distances to the nearest marking using breadth-first search on the ICFG.

\textbf{Loop-Exit Path Prioritization.}
Cyclic regions (blue circle at \circled{FF0000}{4} in Figure~\ref{fig:overview-architecture}) identified during graph collapse require specific prioritization to prevent \textsc{KLEE} from being trapped in a small portion of the code. In these regions, \tech{} distinguishes between cyclic regions that contain LLM-generated markings and those that do not. If the current cyclic region contains markings, remaining inside the region can still be useful, so \tech{} continues to prioritize paths that move closer to those markings. If the current cyclic region contains no marking, \tech{} prioritizes outgoing edges that leave the region, avoiding redundant loop-heavy exploration that is unlikely to expose security violations. It ensures that the exploration does not stall within loops.

\textbf{LLM-Guided Path Prioritization.}
LLM-generated markings are associated with their corresponding nodes in the graph (\circled{FF0000}{4} in Figure~\ref{fig:overview-architecture}). For ordinary non-cyclic nodes, path prioritization directly follows the computed distance to these markings. This guidance allows the engine to focus on paths leading to potential vulnerabilities identified by an LLM.

\subsection{Custom Searcher}
Based on Section~\ref{sec:phase2}, our path prioritization searcher follows four ordered steps, as shown in \circled{FF0000}{5} of Figure~\ref{fig:overview-architecture}. The four steps are applied sequentially: (C1) prioritize paths with a shorter distance to LLM-generated markings, following the distance-based guidance of SDSE~\citep{ma2011directed}. Once a marked node is reached, its marking is removed, and the node is treated as an ordinary node. This does not affect any remaining markings reachable through that node; (C2) prioritize paths that exit the cyclic region without any LLM-generated marking, which is the key design that couples loop handling with LLM-generated markings; (C3) prefer paths that have greater basic-block depth in the ICFG as an auxiliary rule; and (C4) return to random selection after executing 35 million instructions without covering any new basic block. We set this threshold because such long coverage stalls rarely occurred in our preliminary experiments, so that C4 serves only as a fallback when the searcher may continue prioritizing paths toward unreachable markings.

\section{Evaluation}

To comprehensively assess the effectiveness of \tech{}, we formulate the following four Research Questions (RQs).

\begin{description}[style=multiline, leftmargin=8mm, topsep=0pt, partopsep=0pt]
\item[RQ1] \textbf{Code Coverage:} \textit{How effective is \tech{} in improving the code coverage of \textsc{KLEE?}} This RQ examines the effectiveness of \tech{} in guiding \textsc{KLEE} to explore uncovered code that is difficult to reach in a given program within a practical time limit~\citep{klee2019engine,busse2024ssle,zhu2025optse}. This question is essential because the path explosion problem prevents \textsc{KLEE} from exploring all possible execution paths.
\end{description}

\begin{description}[style=multiline, leftmargin=8mm, topsep=0pt, partopsep=0pt]
\item[RQ2] \textbf{Violation Discovery:} \textit{How effective is \tech{} in revealing security violations?} This RQ aims to evaluate the effectiveness of \tech{} in guiding \textsc{KLEE} to expose security violations of a given program within a practical time. Because critical security violations may remain hidden in hard-to-reach execution paths, it is important to investigate whether \textsc{KLEE} can mathematically trigger these exploit conditions before exhausting its time budget~\citep{yi2024cbc,tu2024vital}.
\end{description}

\begin{description}[style=multiline, leftmargin=8mm, topsep=0pt, partopsep=0pt]
\item[RQ3] \textbf{Model Sensitivity:} \textit{How do different LLM families and parameter scales used in \tech{} influence its effectiveness?} This RQ examines the effectiveness of \tech{} under different LLM families and parameter scales in practical settings.
Since prior vulnerability detection studies~\citep{lin2025comparative,ding2025vulnerability,sun2024llm4vuln} showed that LLM-based detection is highly sensitive to model choice and trained parameters, it is important to examine whether different models can provide comparably useful vulnerability signals to guide symbolic execution.
\end{description}

\begin{description}[style=multiline, leftmargin=8mm, topsep=0pt, partopsep=0pt]
\item[RQ4] \textbf{Ablation Study:} \textit{How does each component contribute to the effectiveness of \tech{}?} This RQ isolates the effect of the main design choices in \tech{} through ablation.
Because \tech{} is an LLM-guided vulnerability-oriented directed symbolic execution approach, it is important to distinguish how each component in Figure~\ref{fig:overview-architecture} (searcher, internal components of \tech{}, marking sources, and prompt) contributes to overall effectiveness and whether the observed gains are based on one alone or on their combination.
\end{description}

\subsection{Experimental Setup}
\label{sec:setup}

\textbf{Baseline Exploration Strategies.}
Consistent with the baselines used in the evaluation design for the state-of-the-art path-prioritization approach, \textsc{Empc}~\citep{empc2025}, we compare \tech{} with 13 baseline exploration strategies, including nine built-in \textsc{KLEE} strategies~\citep{cadar2008klee,klee}, \path{bfs}, \path{dfs}, \path{random-path}, \path{random-state}, \path{nurs:covnew}, \path{nurs:md2u}, \path{nurs:rp}, \path{nurs:cpicnt}, and \path{nurs:qc}; excluding \path{nurs:depth} and \path{nurs:icnt}, following \textsc{Empc}'s evaluation design.
Moreover, we compare four representative path-prioritization approaches from previous studies: \textsc{SGS}, \textsc{Learch}, \textsc{CGS}, and \textsc{Empc}. 




\begin{itemize}
    \item \textsc{SGS}~\citep{li2013steering} prioritizes paths whose subpaths have been explored less frequently, thereby steering symbolic execution toward less-traveled regions of the program. Following its original study, we use four independent instances with subpath lengths 1, 2, 4, and 8, each allocated one quarter of the total time budget. Following \textsc{Empc}'s evaluation~\citep{empc2025}, we port \textsc{SGS} to \textsc{KLEE} 3.1 and LLVM 13.0.1.
    \item \textsc{Learch}~\citep{he2021learch} is a machine-learning-based strategy that selects promising paths under path explosion. It uses pre-trained feedforward models built from features extracted from symbolic-execution behavior and leverages existing heuristics during training-data construction. In our evaluation, we directly use the pre-trained models released in their replication package. Following \textsc{Empc}'s evaluation, we use \textsc{KLEE} 2.1 and LLVM 6.0, as \textsc{Learch} relies on a heavy Python binding built on an earlier \textsc{KLEE}/LLVM environment. 
    \item \textsc{CGS}~\citep{sun2024cgs} is a concrete-constraint-guided strategy designed to improve coverage by prioritizing paths that are likely to traverse partially covered concrete branches. Its guidance is derived from concrete constraints and data dependence and, therefore, differs from purely structural or historical exploration heuristics. We follow \textsc{Empc} and use \textsc{KLEE} 3.0-pre and LLVM 11.0, as \textsc{CGS}'s experimental environment involves substantial modifications to an earlier \textsc{KLEE}.
    \item \textsc{Empc}~\citep{empc2025} reduces redundant exploration through graph-level analysis. It computes representative paths from multiple minimum path covers and uses them to guide symbolic execution toward a smaller but structurally informative subset of the path space. Consistent with its implementation, we use \textsc{KLEE} 3.1 and LLVM 13.0.1 for \textsc{Empc}.
\end{itemize}

\textbf{Experimented Large Language Models.}
To assess the effectiveness of \tech{}, we select four coder models and three larger general-purpose models that are used to identify potentially vulnerable locations (Section \ref{sec:phase1}). For coder models, we select two families of coder models that have already appeared in recent vulnerability-detection evaluations and security-oriented code studies~\citep{bruni2025forge_gpt4o,bae2024gpt4o_vuln,siddiq2025security_comprehension,qin2025reasoning_vuln_eval}: \textit{DeepSeek-Coder}, and \textit{Qwen2.5-Coder}. To further examine how model size affects effectiveness, we include two model sizes from each
coder model family: \textit{6.7b} and \textit{33b} for \textit{DeepSeek-Coder}, and \textit{7b} and \textit{32b} for \textit{Qwen2.5-Coder}.  For larger general-purpose models, we select two models that are widely utilized in vulnerability detection and security~\citep{bruni2025forge_gpt4o,bae2024gpt4o_vuln,siddiq2025security_comprehension}: \textit{Gemini 2.5 Flash} and \textit{GPT-4o}. Additionally, given its extensive adoption in security research~\citep{qin2025reasoning_vuln_eval}, we also include the latest cloud-based version of \textit{DeepSeek}: \textit{DeepSeek-V4-Flash}. To ensure consistency, we disable the thinking feature for \textit{Gemini 2.5 Flash} and \textit{DeepSeek-V4-Flash} according to their official documentation, as additional reasoning processes may alter the model's response to the same prompt~\citep{wei2022cot, wang2023selfconsistency}.

For the four local models, we use the official \textsc{Ollama} versions and limit the context size to 12,288 tokens due to computational resource limitations. We retain their original quantization formats: \textit{DeepSeek-Coder-6.7B} (ce298d984115) and \textit{DeepSeek-Coder-33B} (acec7c0b0fd9) use Q4\_0 quantization, whereas \textit{Qwen2.5-Coder-7B} (dae161e27b0e) and \textit{Qwen2.5-Coder-32B} (b92d6a0bd47e) use Q4\_K\_M quantization. All four local models are loaded and run entirely on a GPU without CPU offloading. For the three cloud models, we use \textit{GPT-4o} through OpenAI, \textit{Gemini 2.5 Flash} through OpenRouter, and \textit{DeepSeek-V4-Flash} through DeepSeek. Their exact model identifiers are \texttt{gpt-4o-2024-08-06}, \texttt{google/gemini-2.5-flash}, and \texttt{deepseek-v4-flash}, respectively. We report more details in our replication package.

For all experimented LLMs, we set the temperature to 0 to ensure greedy decoding to maximize reproducibility~\citep{wei2022cot, wang2023selfconsistency}, while acknowledging that modern LLM inference can still exhibit non-determinism in practice.

\textbf{Benchmark Programs.}
To ensure a fair comparison, we evaluate \tech{} using the same suite of 12 real-world open-source programs as used in \textsc{Empc}. These programs are widely used in fuzzing and symbolic execution techniques~\citep{boehme2017aflgo,busse2020running,he2021learch,kapus2020pending,li2021unifuzz,metzman2021fuzzbench,sun2024cgs}. Table~\ref{tab:bench-details} presents the details of 12 benchmark programs and their exact versions used in our experiments. All \textsc{KLEE} arguments are kept consistent with \textsc{Empc}.

\textbf{Experimental Environment.}
Experiments were conducted on a server with an AMD 3990X (64C/128T), NVIDIA RTX 3090 Graphics Card, 256 GB RAM, and Ubuntu 24.04. Each \textsc{KLEE} instance was limited to 1 core, 16 GB RAM, and 10 hours. Due to resource constraints, we used 16 GB instead of the 32 GB used by~\citet{empc2025} (\textsc{Empc}), though this still exceeds the memory limit in~\citet{he2021learch} (\textsc{Learch}). The NVIDIA RTX 3090 was used to generate markings with local coder models served on Ollama. All experiments ran in Docker containers (v29.1.2).

\begin{table*}[t]
\centering
\caption{Details of the 12 benchmark programs for evaluation.}
\label{tab:bench-details}
\begin{tabular}{@{}llllrrl@{}}
\toprule
\textbf{Project} & \textbf{Program} & \textbf{Ver.} & \textbf{Type} &
\makecell[r]{\textbf{\# Basic}\\\textbf{Blocks}} &
\makecell[r]{\textbf{\# Code}\\\textbf{Lines}} &
\textbf{Created} \\
\midrule
GNU bc       & bc        & 1.08.2  & Calc.   &   2,438 &  11,612 & 1991       \\
GNU ncurses  & tic       & 6.5     & Text    &  39,992 &  29,955 & 1993       \\
GNU bison    & bison     & 3.8.2   & Text    & 124,115 &  81,395 & 1985       \\
GNU binutils & readelf   & 2.45    & Binary  & 146,092 & 111,229 & 1988-07-11 \\
GNU binutils & strip-new & 2.45    & Binary  & 333,714 & 189,043 & 1988-07-11 \\
GNU make     & make      & 4.4.1   & Text    &  39,211 &  31,102 & 1988       \\
NASM         & nasm      & 3.01rc7 & Binary  &  70,386 &  70,553 & 1996       \\
libtiff      & tiffinfo  & 4.7.1   & Image   &  61,882 &  31,842 & 1988       \\
JasPer       & jasper    & 4.2.8   & Image   &  72,762 &  40,372 & 1999-09    \\
Little CMS   & transicc  & 2.17    & Image   &  62,208 &  37,194 & 1998       \\
FLVMeta      & flvmeta   & 1.2.2   & Video   &  27,460 &  11,160 & 2007-09-19 \\
curl         & curl      & 8.16.0  & Net.    & 174,461 & 107,292 & 1998-03-20 \\
\bottomrule
\end{tabular}
\end{table*}

\subsection{RQ1: Code Coverage}
\label{sec:rq1_results}

\begin{figure*}[htbp]
  \centering
  \includegraphics[width=\textwidth]{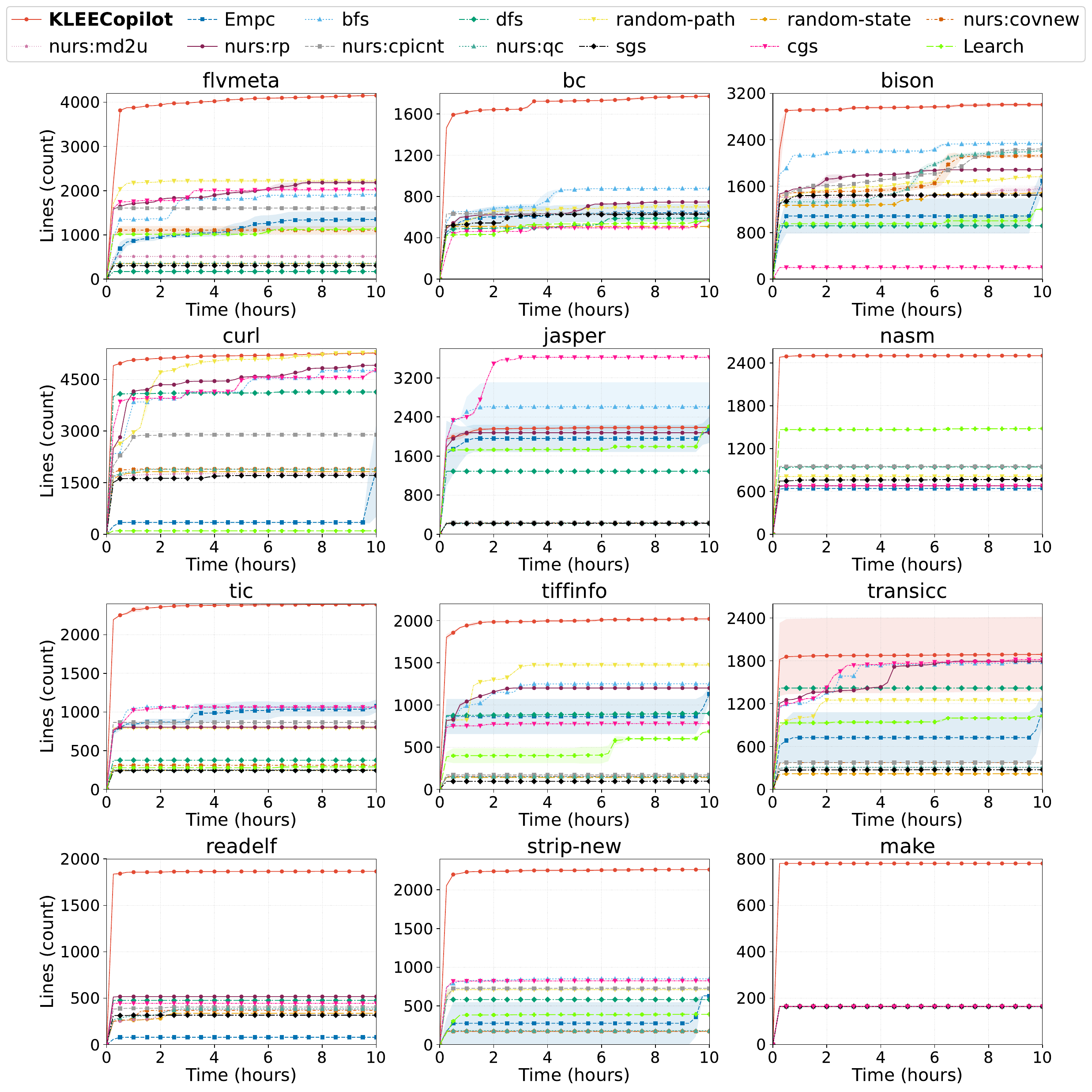}
  \caption{Line coverage over time during ten hours of symbolic exploration}
  \label{fig:rq1_line_cov}
\end{figure*}

\noindent\textbf{Approach.}
In this RQ, we investigate the effectiveness of \tech{} in guiding \textsc{KLEE} in exploring code that is difficult to reach within a practical time limit. To do that, we systematically compare \tech{} powered by \textit{DeepSeek-Coder-33B} (this model is deliberately selected for comparison with other baselines; the performance of the remaining LLMs is presented in Section~\ref{sec:rq3_results}) with 13 baselines (exploration strategies) in 12 benchmark programs in terms of cumulative \textit{line coverage} and \textit{basic-block coverage} during symbolic exploration. 

To evaluate the effectiveness of exploration strategies, we deliberately measure both coverage metrics during \textsc{KLEE}'s execution rather than from replayed test cases. This choice matches the goal of RQ1, which is to evaluate how exploration strategies guide \textsc{KLEE} to explore code that is difficult to reach, rather than to evaluate the coverage of KLEE-generated test cases. In symbolic execution, exploration heuristics are explicitly designed to improve exploration coverage under a limited budget~\citep{cha2018heuristics}. In fuzzing and broader security testing evaluation, the standard practice is to also compare techniques by the coverage achieved during the testing itself, as reflected in widely used benchmarking platforms, e.g., \textsc{FuzzBench}~\citep{metzman2021fuzzbench} and \textsc{UNIFUZZ}~\citep{li2021unifuzz}. The coverage during replayed test cases remains useful for assessing the generated test suite, but it also depends on downstream replay, concretization, build, and sanitizer effects. We therefore use code coverage during \textsc{KLEE}'s execution as the RQ1 metric and apply the same measurement method uniformly to all baselines. We repeat each experiment five times, allowing each KLEE instance to run for ten hours.

\begin{figure*}[htbp]
  \centering
  \includegraphics[width=\textwidth]{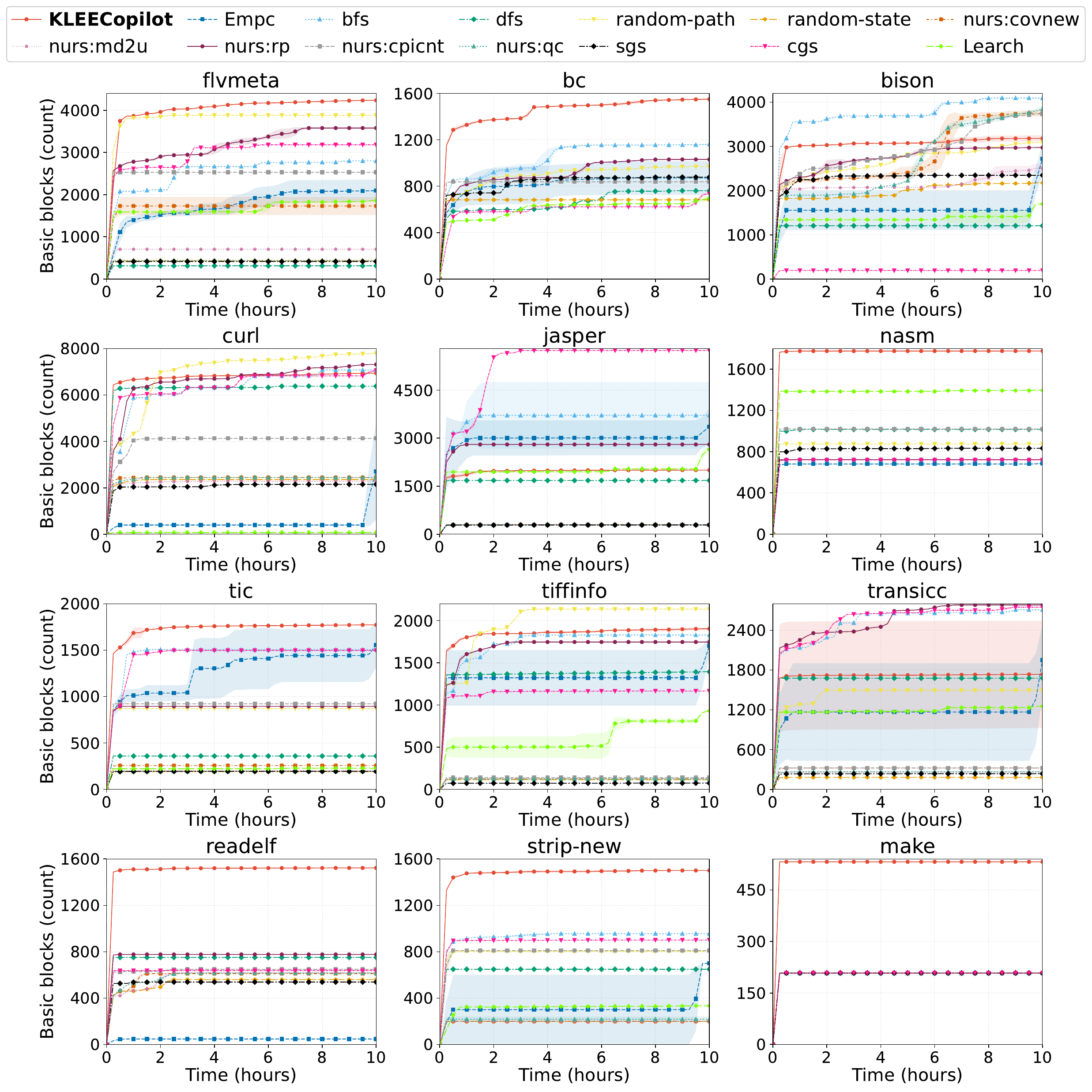}
  \caption{Basic-block coverage over time during ten hours of symbolic exploration}
  \label{fig:rq1_bb_cov}
\end{figure*}

\smallskip\noindent\textbf{Results.}
Figure~\ref{fig:rq1_line_cov} and Figure~\ref{fig:rq1_bb_cov} present, respectively, the evolution of line coverage and basic-block coverage over time for the 14 strategies under comparison. Each curve represents the average cumulative coverage across five runs, while the shaded region indicates variability across runs.

\textbf{Line Coverage.} Figure~\ref{fig:rq1_line_cov} shows that \tech{} achieves the strongest line coverage on 11 of 12 benchmarks. For these 11 benchmarks, \tech{} achieves up to twice (e.g., \path{flvmeta}), three times (e.g., \path{readelf}), or even more than four times (e.g., \path{make}) the line coverage of the second-best exploration strategy. The only benchmark where \tech{} performs worse is \texttt{jasper}, where it is outperformed by \textsc{Empc} and \textsc{CGS}. This is noteworthy because \tech{} is not designed as a coverage-maximizing strategy. Instead, it directs exploration toward LLM-generated markings and simplifies redundant loop-heavy expansions. Our results show that this guidance did not sacrifice code coverage, although it was designed for more focused bug discovery. Furthermore, despite the non-deterministic behavior of LLMs, \tech{} still consistently achieves stable coverage on 11 of the 12 benchmarks, with observable variation only in \path{transicc}. 

\textbf{Basic-Block Coverage.} 
Figure~\ref{fig:rq1_bb_cov} presents a weaker, but still generally consistent trend in basic-block coverage as line coverage. At the 10-hour point, \tech{} attains the highest basic-block coverage on seven of the 12 benchmarks.
The best-performing cases still exhibit substantial improvements, such as \texttt{flvmeta} (+355.8 basic blocks over the second-best baseline), \texttt{make} (+324.0), and \texttt{nasm} (+380.8). 
The clearly weak cases are \texttt{transicc} (\(-37.61\%\)), \texttt{bison} (\(-22.33\%\)), \texttt{curl} (\(-10.92\%\)), \texttt{tiffinfo} (\(-10.88\%\)), and in particular \texttt{jasper} (\(-65.11\%\)). Thus, unlike the line coverage metric, the basic-block metric reveals a mixed pattern: several strong wins and several pronounced losses concentrated on structurally challenging benchmark programs.

\begin{summarybox}
Although \tech{} is not designed as a coverage-maximizing strategy, it still outperformed other baselines by improving \textsc{KLEE}'s coverage substantially within a practical time budget. It yields 125.82\% higher line coverage and 42.24\% higher basic-block coverage than the latest baseline, \textsc{Empc}.
\end{summarybox}

\subsection{RQ2: Violation Discovery}
\label{sec:rq2_results}

\textbf{Approach.}
This RQ evaluates how effectively \tech{} can uncover security violations by replaying the test cases generated when \tech{} is guiding \textsc{KLEE}. To this end, we systematically compare it, using \textit{DeepSeek-Coder-33B} (same as RQ1), against 13 baseline strategies in 11 benchmark programs, measuring both the
total \textit{number of violations} and the \textit{number of unique violations}. 

For this RQ, we repeat the experiment five times (following \textsc{Empc}~\citep{empc2025}), replaying test cases (generated by all strategies when guiding \textsc{KLEE}) on programs instrumented with Undefined Behavior Sanitizer (UBSan) and Address Sanitizer (ASan) to uncover additional security violations five times. In particular, the replay binaries are compiled with \path{-fsanitize=undefined,address,integer,bounds}. Because \texttt{bison} is incompatible with UBSan, it is excluded from the original set of 12 benchmark programs. We obtain the total number of violations by aggregating the outputs from replaying the test cases produced by \textsc{KLEE} and then recording the maximum value across the five replays of each test case. For unique violations, we deduplicate the discovered violations using the CASR-style approach~\citep{savidov2021casrcluster}, which has been adopted in fuzzing evaluations~\citep{wu2025wildsync,yegorov2024pythonfuzzing,leonelli2025twinfuzz}, across the five experiments and the five replays of each test case. Furthermore, we also visualize the overlaps of discovered violations among strategies using the Venn diagram, showing how \tech{} is able to successfully discover particular security violations that all existing baseline strategies completely miss.

\begin{figure}[t]
  \centering
  \includegraphics[width=0.6\columnwidth]{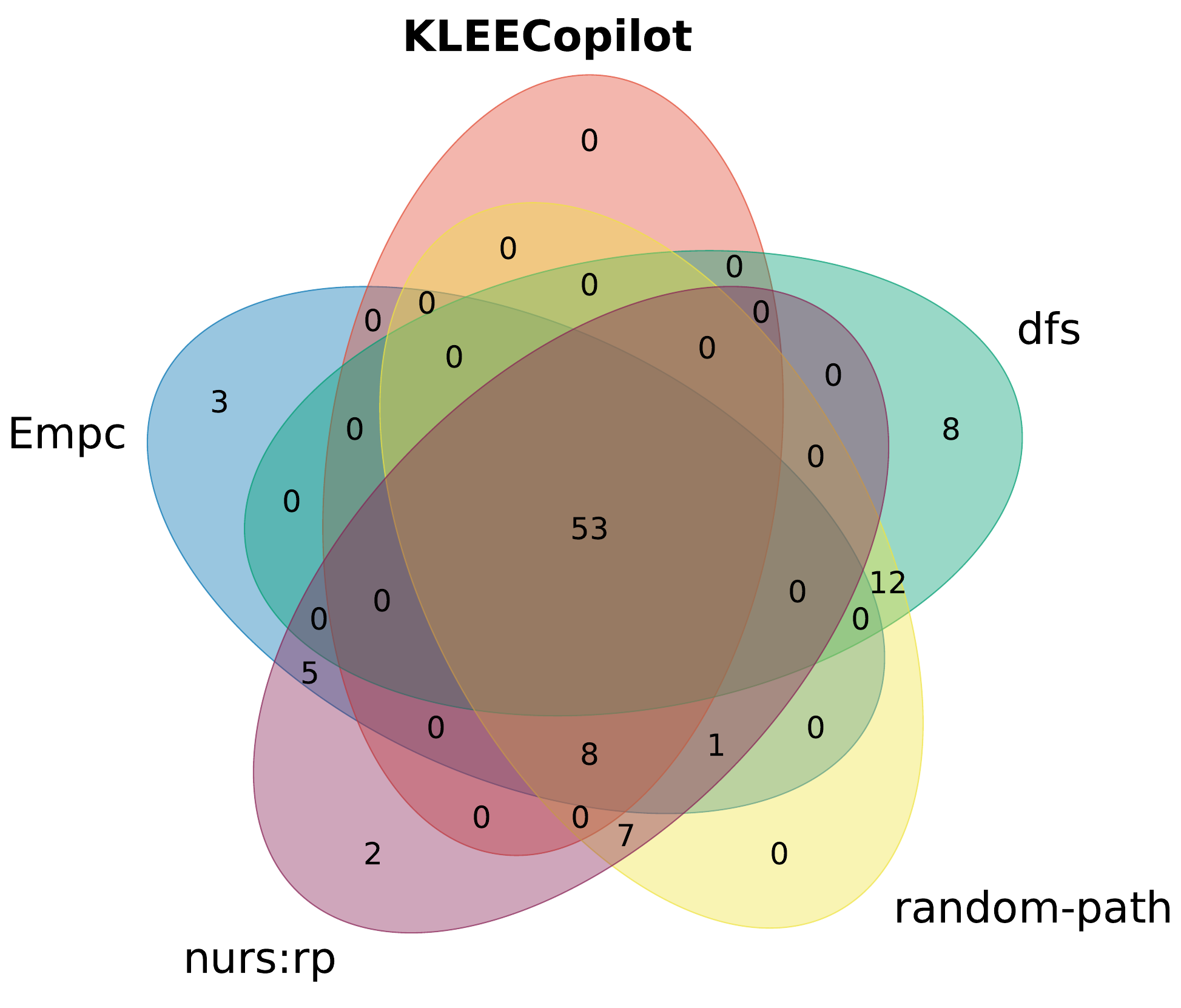}
  \caption{Venn diagrams of unique violations discovered by the top-5 ranking strategies}
  \label{fig:crash-overlap}
\end{figure}

\smallskip\noindent\textbf{Results.}
Table~\ref{tab:rq2-violations} reports both total and unique violation counts as well as the number of benchmark programs with at least one discovered violation by \tech{} along with 13 baseline strategies on 11 benchmarks. Figure~\ref{fig:crash-overlap} illustrates the overlaps of the discovered violations among the top-5 ranking strategies.

\begin{table*}[t]
\centering
\caption{The total number of violations and number of unique violations discovered by \tech{} and all baselines on 11 benchmarks.}
\label{tab:rq2-violations}

\begin{threeparttable}
\begin{tabular}{lrrr}
\toprule
\textbf{Strategy} &
\textbf{\# Violations} &
\textbf{\# Unique Violations} &
\textbf{\# Benchmarks}\tnote{*} \\
\midrule
\textbf{KLEECopilot} & \textbf{1,335} & \textbf{87} & \textbf{9}\\
random-path  & 1,010 & 81 & 9\\
nurs:rp      & 919  & 76 & 9\\
dfs          & 775  & 73 & 7\\
Empc         & 721  & 70 & 9\\
CGS          & 1,010 & 63 & 9\\
bfs          & 896  & 61 & 8\\
nurs:qc      & 480  & 57 & 6\\
nurs:covnew  & 509  & 55 & 5\\
nurs:cpicnt  & 537  & 54 & 6\\
SGS          & 540  & 54 & 6\\
random-state & 440  & 39 & 4\\
nurs:md2u    & 445  & 38 & 4\\
Learch       & 70   & 9  & 5\\
\bottomrule
\end{tabular}

\begin{tablenotes}
\footnotesize
\item[*] \# Benchmarks indicates the number of benchmark programs that have at least one discovered violation for each corresponding strategy.
\end{tablenotes}
\end{threeparttable}
\end{table*}

\textbf{Total Violations.} Table~\ref{tab:rq2-violations} shows that the five best-performing strategies for the total number of violations are \tech{} (1,335), followed by four built-in strategies of \textsc{KLEE}: \path{random-path} (1,010), \textsc{CGS} (1,010), \path{nurs:rp} (919), and \path{bfs} (896). Compared with the second-best baselines, \path{random-path} and \textsc{CGS}, \tech{} uncovers 325 additional violations, representing a 32.2\% improvement. The state-of-the-art path-prioritization approach, \textsc{Empc}, ranks only seventh (721).  \tech{} discovers 614 more violations in total, which corresponds to an 85.2\% increase. In addition, \tech{} also ties for the highest benchmark coverage, uncovering at least one violation in nine out of 11 benchmarks, comparable to the effectiveness of \path{random-path} and \path{nurs:rp}.

\textbf{Unique Violations.} Table~\ref{tab:rq2-violations} shows that the top five strategies in terms of unique violations are \tech{} (87), \path{random-path} (81), \path{nurs:rp} (76), \path{dfs} (73), and \textsc{Empc} (70). Compared to the second-best strategy, \tech{} finds six additional unique violations (a 7.4\% increase), and compared to \textsc{Empc}, it discovers 17 more, corresponding to a 24.3\% improvement.

Figure~\ref{fig:crash-overlap} shows that the top-5 strategies share a large common core of 53 unique violations. The main differences lie in the partially overlapping and exclusive regions. \tech{} contributes the largest exclusive region, with 7 violations, slightly more than \path{dfs}'s 6. These exclusive violations are marking-driven: they are exposed on execution paths prioritized by distances computed from LLM-generated markings.

Although \path{random-path} and \path{nurs:rp} rank second and third by the number of unique violations, they contribute few exclusive violations: \path{random-path} contributes none and \path{nurs:rp} contributes only two. This suggests that their strength mainly comes from covering the shared violation space. However, the seven violations shared only by \path{random-path} and \path{nurs:rp} show the value of randomized selection. \path{random-path} explicitly samples a path through the execution tree and selects a leaf state, while \path{nurs:rp} is built on \path{random-path} and prefers shallower paths. These strategies can preserve states whose selection is not biased by graph cover (such as in \textsc{Empc}) or by data dependence (such as in \textsc{CGS}). A representative example is \textit{flvmeta}, where several violations come from combinations of non-relevant individual branch conditions.

\path{dfs}, which contributes the second-largest unique region, reflects another bias. \path{dfs} selects recently generated states from the state queue, which often keeps exploration in a deeper path. Thus, the six violations do not indicate any specific guidance; rather, they show that \path{dfs} spends more budget on deeper states. These cases have clear characteristics. For example, in \path{nasm}, the \texttt{response-file} handling logic contains recursive call processing, and this region is not covered by LLM-generated markings. According to our loop-exit prioritization, \tech{} does not continue prioritizing such unmarked loop regions, whereas \path{dfs} can still steer on these deeper states.

Nevertheless, \tech{}, \path{dfs}, and \path{random-path} share 12 violations that are missed by \textsc{Empc} and \path{nurs:rp}. These 12 violations are not in very shallow execution regions. In contrast, \path{nurs:rp} tends to select shallower states because of its depth-based weight, and \textsc{Empc} needs to make decisions based on graph coverage. \tech{}, \path{dfs}, and \path{random-path} are different: \path{dfs} selects recently generated states, which are usually deeper; \path{random-path} selects a leaf state by sampling a path in the execution tree; and \tech{} prioritizes states according to  LLM-generated markings. Therefore, these three strategies are more likely to continue exploring already formed deeper paths, which explains why they share these 12 violations.

\textbf{Confirmation of Discovered Violations.}
Because our experiments use the latest versions of real-world programs, we have reported the 18 identified security violations to the corresponding developers. The first author of this paper, who has eight years of industrial experience in software security, reviewed these violations from a total of 87 unique violations discovered by \tech{} to avoid reporting noise (e.g., an unsigned integer overflow cannot trigger subsequent memory corruptions). We subsequently received confirmation of the existence of 12 violations, while nine have been fixed in the corresponding benchmark programs. In our replication package, we include the bug reports and pull requests associated with the identified violations.

Table~\ref{tab:rq2-analysis} presents the distributions of the 18 reported violations and of all unique violations discovered by \tech{}, where a violation confirmed by developers is regarded as a security issue and otherwise as undefined behavior detected by UBSan.

\begin{table*}[t]
\centering
\caption{Distributions of the unique, reported, and fixed violations.}
\label{tab:rq2-analysis}
\resizebox{\columnwidth}{!}{
\begin{tabular}{@{}lrrr@{}}
\toprule
\textbf{Category} &
\textbf{\# Unique Violations} &
\textbf{\# Reported Violations} &
\textbf{\# Fixed Violations} \\
\midrule
Security issue     & 12 & 12 & 9 \\
Undefined behavior & 75 & 6  & 0 \\
\midrule
Total              & 87 & 18 & 9 \\
\bottomrule
\end{tabular}
}
\end{table*}

The consequences of undefined behavior vary across contexts. Accordingly, developers held different views on the same class of undefined behavior: a similar unsigned integer overflow was fixed in \texttt{flvmeta} but not fixed in \texttt{transicc}. To avoid disturbing the open-source community, we therefore reported only representative instances of each class.

\begin{summarybox}
\tech{} achieves the best security violation discovery results among all baselines, reaching 1,335 total security violations and 87 unique violations, which exceed the second-best baselines by 325 (+32.2\%) and 6 (+7.4\%), respectively, and exceed \textsc{Empc} by 614 (+85.2\%) and 17 (+24.3\%), respectively.
\end{summarybox}

\subsection{RQ3: Model Sensitivity}
\label{sec:rq3_results}

\textbf{Approach.}
This RQ investigates the sensitivity of \tech{} to the choice of the LLM used. Prior studies have shown that different LLMs can show different effectiveness in vulnerability detection~\citep{lin2025comparative,ding2025vulnerability,sun2024llm4vuln,ahmed2025secvuleval}. Since different LLMs may also produce different vulnerable markings for the same benchmark program, an important question is how these marking differences affect the performance of coverage and security violations in \tech{}. We keep all other experimental settings unchanged and only change the LLM used to generate security violation markings. The evaluated models are listed in Section~\ref{sec:setup}. We select \textit{DeepSeek-Coder-33B} as the reference model, as it showed the highest performance in violation discovery and competitive code coverage. Furthermore, we compare the results against three baselines: (i) \textsc{CGS}, which demonstrated competitive code coverage on specific benchmark programs (e.g., \textit{jasper}) in RQ1; (ii) \path{random-path}, which serves as the second-best baseline for unique security violation discovery in RQ2; and (iii) \textsc{Empc}, the state-of-the-art path-prioritization approach.

\begin{table}[t]
\centering
\caption{The overall effectiveness of seven experimented LLMs employed in \tech{} and three baseline strategies.}
\label{tab:llm-coverage-compare}

\begin{tabular}{@{}lcrrrr@{}}
\toprule
\textbf{Model / Strategy} & \textbf{\#UV} &
\makecell[c]{\textbf{BB Median}\\$\Delta$ (\%)} &
\makecell[c]{\textbf{BB Avg.}\\$\Delta$ (\%)} &
\makecell[c]{\textbf{Line Median}\\$\Delta$ (\%)} &
\makecell[c]{\textbf{Line Avg.}\\$\Delta$ (\%)} \\
\midrule

DeepSeek-Coder-33B & 87
  & \makecell[tr]{--\\--}
  & \makecell[tr]{--\\--}
  & \makecell[tr]{--\\--}
  & \makecell[tr]{--\\--} \\

DeepSeek-Coder-6.7B & 87
  & \makecell[tr]{-2.00\\(-0.11\%)}
  & \makecell[tr]{-23.98\\(-1.00\%)}
  & \makecell[tr]{-1.10\\(-0.05\%)}
  & \makecell[tr]{-13.15\\(-0.52\%)} \\

GPT-4o & 87
  & \makecell[tr]{-36.30\\(-2.04\%)}
  & \makecell[tr]{\textbf{+31.77}\\(\textbf{+1.33\%})}
  & \makecell[tr]{-26.20\\(-1.18\%)}
  & \makecell[tr]{\textbf{+15.38}\\(\textbf{+0.61\%})} \\

DeepSeek-V4-Flash & 87
  & \makecell[tr]{-31.90\\(-1.80\%)}
  & \makecell[tr]{-54.27\\(-2.27\%)}
  & \makecell[tr]{-19.80\\(-0.89\%)}
  & \makecell[tr]{-38.33\\(-1.53\%)} \\

random-path\tnote{*} & 81
  & \makecell[tr]{-340.50\\(-19.18\%)}
  & \makecell[tr]{-301.14\\(-11.98\%)}
  & \makecell[tr]{-1158.20\\(-49.76\%)}
  & \makecell[tr]{-1087.52\\(-41.75\%)} \\

Empc & 70
  & \makecell[tr]{-567.40\\(-31.96\%)}
  & \makecell[tr]{-671.62\\(-28.12\%)}
  & \makecell[tr]{-1309.90\\(-58.91\%)}
  & \makecell[tr]{-1361.80\\(-54.29\%)} \\

CGS & 63
  & \makecell[tr]{-669.10\\(-37.69\%)}
  & \makecell[tr]{-321.90\\(-13.48\%)}
  & \makecell[tr]{-1285.40\\(-57.80\%)}
  & \makecell[tr]{-1095.43\\(-43.67\%)} \\

Gemini 2.5 Flash & 60
  & \makecell[tr]{-6.60\\(-0.37\%)}
  & \makecell[tr]{\textbf{+6.50}\\(\textbf{+0.27\%})}
  & \makecell[tr]{-4.20\\(-0.19\%)}
  & \makecell[tr]{-3.92\\(-0.16\%)} \\

Qwen2.5-Coder-7B & 60
  & \makecell[tr]{-9.90\\(-0.56\%)}
  & \makecell[tr]{\textbf{+36.98}\\(\textbf{+1.55\%})}
  & \makecell[tr]{-3.10\\(-0.14\%)}
  & \makecell[tr]{\textbf{+18.70}\\(\textbf{+0.75\%})} \\

Qwen2.5-Coder-32B & 54
  & \makecell[tr]{\textbf{+0.80}\\(\textbf{+0.05\%})}
  & \makecell[tr]{\textbf{+5.73}\\(\textbf{+0.24\%})}
  & \makecell[tr]{\textbf{+2.20}\\(\textbf{+0.10\%})}
  & \makecell[tr]{-0.10\\(0.00\%)} \\

\bottomrule
\end{tabular}

\begin{tablenotes}
\footnotesize
\item Each coverage cell shows the absolute difference above its percentage difference.
\item[*] For \path{random-path}, both medians and averages are based on
10 benchmarks due to \textsc{KLEE} state exhaustion in \textit{jasper}
and \textit{readelf} before coverage collection; other strategies use
all 12 benchmarks. UV: unique security violation; BB: basic-block.
\end{tablenotes}

\end{table}


\smallskip\noindent\textbf{Results.}
Table~\ref{tab:llm-coverage-compare} presents the results of model sensitivity among seven experimented LLMs and three selected baseline strategies. All coverage columns report median and average deltas across all benchmark programs relative to the reference model, with bold indicating positive deltas relative to \tech{} with \textit{DeepSeek-Coder-33B}.

\textbf{Comparison of Security Violation Discovery.} 
Table~\ref{tab:llm-coverage-compare} shows that model choice affects the number of unique security violations discovered. The three DeepSeek models, including \textit{DeepSeek-Coder-33B}, \textit{DeepSeek-Coder-6.7B}, and \textit{DeepSeek-V4-Flash}, each discover 87 unique violations. \textit{GPT-4o} achieves the same result, whereas \textit{Gemini 2.5 Flash}, \textit{Qwen2.5-Coder-7B}, and \textit{Qwen2.5-Coder-32B} discover only 60, 60, and 54 unique violations, respectively. These results suggest that security violation discovery is more closely related to model characteristics than to model parameter scale. The consistent results among the three DeepSeek models may reflect shared characteristics in their training and development, such as partially overlapping training data, training objectives, or training methods. However, because these factors are not independently controlled in our experiments, we cannot attribute their performance to any specific factor. Moreover, because of the fundamental limitations of \textsc{KLEE}, some markings cannot be reached. Future work should account for reachability under symbolic execution engine limitations when generating vulnerability markings, since unreachable markings will lead to unnecessary exploration.

\textbf{Comparison of Code Coverage.} The LLM variants used by \tech{} consistently outperform the three baseline strategies in code coverage. Compared with \textit{DeepSeek-Coder-33B}, the baseline strategies reduce median basic-block coverage by 19.18--37.69\% and median line coverage by 49.76--58.91\%. Among the other open source models, \textit{Qwen2.5-Coder-32B} achieves the highest median line coverage, improving it by 0.10\%, while \textit{Qwen2.5-Coder-7B} achieves the highest average basic block coverage, improving it by 1.55\%. However, these slight coverage improvements do not lead to more discovered unique security violations: the two Qwen models discover only 54--60, compared with 87 for \textit{DeepSeek-Coder-33B}. This again indicates that higher coverage alone does not necessarily improve security violation discovery due to the limitation of \textsc{KLEE}. Moreover, prior work has shown that some feasible paths cannot be effectively explored because of complex path constraints~\citep{baldoni2018survey}.

\begin{summarybox}
Our results show that model choice affects both coverage and the discovery of security violations. \textit{DeepSeek-Coder-33B} achieves a joint-best security-violation discovery result, with 87 unique violations, together with \textit{DeepSeek-Coder-6.7B}, \textit{GPT-4o}, and \textit{DeepSeek-V4-Flash}. In contrast, \textit{Qwen2.5-Coder-32B} shows positive coverage deltas across three coverage metrics, but discovers only 54 unique violations. This suggests that, under directed symbolic execution, coverage improvement alone is insufficient. The quality of vulnerability markings is crucial for guiding \textsc{KLEE} toward security violations.
\end{summarybox}

\subsection{RQ4: Ablation Study}
\label{sec:rq4_results}

\textbf{Approach.} This RQ evaluates how the major components in Figure~\ref{fig:overview-architecture} contribute to \tech{}'s ability to find security violations and reach uncovered code of a given program. We conduct an ablation study using the same configurations as in RQ3, selecting \textit{DeepSeek-Coder-33B} as our representative model due to its performance. We organize the ablation study into four groups (searcher, internal components
of \tech{}, marking sources, and prompt).

First, to compare against the directed symbolic execution strategy as discussed in Section~\ref{sec:rw}, we implement the \textbf{\textsc{SDSE}}~\citep{ma2011directed} baseline in the same \textsc{KLEE} version with the same LLM-generated markings (\textit{DeepSeek-Coder-33B}) used by \tech{}. The \textsc{SDSE} variant prioritizes states according to the shortest interprocedural distance to marking locations, without applying \tech{}'s loop-exit prioritization.

Second, we compare \tech{} against two internal algorithm variants:
\begin{itemize}
    \item $v_1$ (\textbf{\tech{} w/o LLM-Guided Path Prioritization}): This variant disables the LLM-generated
marking guidance and its subsequent integration, indicating that 
\textsc{KLEE} relies only on the loop-exit path prioritization, without guidance to potentially vulnerable locations. 
\item $v_2$ (\textbf{\tech{} w/o Loop-Exit Path Prioritization}): This variant disables the loop-exit path prioritization, meaning \textsc{KLEE} only relies on LLM-guided path prioritization.
\end{itemize}

Third, we evaluate whether the effectiveness of \tech{} comes specifically from different marking sources. For this purpose, we keep the same searcher (using also loop-exit path prioritization) as \tech{} but replace the marking source with five alternatives:

\begin{itemize}
    \item \textbf{Random markings}: To show that the improvements in both unique violations and coverage are not coincidental, we randomly mark basic blocks while matching the average function-level marking ratio of markings generated by four local LLMs (8.03\%).
    \item \textbf{Static analysis markings}: Inspired by directed symbolic execution from static analysis~\citep{babic2011sandwich, gerasimov2018directed, busse2022combining}, we use warnings produced by CodeQL~\citep{codeql}, Infer~\citep{infer}, and Semgrep~\citep{semgrep} mapped to basic blocks as markings. To ensure accuracy, CodeQL and Infer require successful compilation, while Semgrep directly scans the codebase.
    \item \textbf{Sink heuristic markings}: To distinguish the code semantic understanding of LLMs from merely marking potentially vulnerable sink functions, we extract C/C++ relevant sink functions from CodeQL, Infer, and Semgrep official rules, then mark call sites at the basic block level to produce markings.
    \item \textbf{Sanitizer heuristic}: To demonstrate that sanitizer instruments cannot substitute for LLMs' security knowledge, we compare LLM-generated markings with a sanitizer heuristic baseline, we mark all basic blocks instrumented by ASan (Address Sanitizer) or UBSan (Undefined Behavior Sanitizer) as markings; these two sanitizers are consistent with security violation types in our LLM prompt (Section~\ref{sec:phase1}).
    \item \textbf{No markings}: To isolate the effect of loop-exit prioritization from LLM-guided path prioritization, we further set all basic block markings to false. This is different from $v_1$ in the internal algorithm ablation study, where the LLM-guided path prioritization component is removed.
\end{itemize}

Fourth, we have both positive and negative prompts in our design (Figure~\ref{fig:prompt}). To distinguish the effects of positive and negative prompts, we use two prompt variants: the \textbf{positive-only variant} uses only positive examples, while the \textbf{negative-only variant} uses only negative examples.

In addition, we show representative baseline strategies from RQ1 and RQ2, including \path{random-path}, Empc, and CGS. All coverage columns report median and average deltas relative to \tech{} with \textit{DeepSeek-Coder-33B}, and positive deltas are highlighted in bold.

\begin{table}[!htbp]
\centering
\caption{Ablation results for \tech{} with searcher design, algorithmic components, marking sources, and prompt sensitivity, and baseline strategies.}
\label{tab:ablation-coverage-compare}

\begin{tabular}{@{}lcrrrr@{}}
\toprule
\textbf{Variant} & \textbf{\#UV} &
\makecell[c]{\textbf{BB Median}\\$\Delta$ (\%)} &
\makecell[c]{\textbf{BB Avg.}\\$\Delta$ (\%)} &
\makecell[c]{\textbf{Line Median}\\$\Delta$ (\%)} &
\makecell[c]{\textbf{Line Avg.}\\$\Delta$ (\%)} \\
\midrule

\multicolumn{6}{@{}l}{\textbf{Searcher}} \\

\tech{} & \textbf{87}
  & \makecell[tr]{--\\--}
  & \makecell[tr]{--\\--}
  & \makecell[tr]{--\\--}
  & \makecell[tr]{--\\--} \\

\textsc{SDSE} & 54
  & \makecell[tr]{-126.90\\(-7.15\%)}
  & \makecell[tr]{-96.77\\(-4.05\%)}
  & \makecell[tr]{-138.60\\(-6.23\%)}
  & \makecell[tr]{-148.52\\(-5.92\%)} \\

\midrule
\multicolumn{6}{@{}l}{\textbf{Internal Components}} \\

$v_1$ & 61
  & \makecell[tr]{\textbf{+7.20}\\(\textbf{+0.41\%})}
  & \makecell[tr]{\textbf{+124.78}\\(\textbf{+5.23\%})}
  & \makecell[tr]{\textbf{+0.20}\\(\textbf{+0.01\%})}
  & \makecell[tr]{\textbf{+82.72}\\(\textbf{+3.30\%})} \\

$v_2$ & 58
  & \makecell[tr]{-106.60\\(-6.01\%)}
  & \makecell[tr]{-105.12\\(-4.40\%)}
  & \makecell[tr]{-80.10\\(-3.60\%)}
  & \makecell[tr]{-90.85\\(-3.62\%)} \\

\midrule
\multicolumn{6}{@{}l}{\textbf{Marking}} \\

Random markings & 61
  & \makecell[tr]{-16.40\\(-0.92\%)}
  & \makecell[tr]{\textbf{+53.32}\\(\textbf{+2.23\%})}
  & \makecell[tr]{-11.40\\(-0.51\%)}
  & \makecell[tr]{\textbf{+28.78}\\(\textbf{+1.15\%})} \\

CodeQL markings & 60
  & \makecell[tr]{-15.60\\(-0.88\%)}
  & \makecell[tr]{-1.80\\(-0.08\%)}
  & \makecell[tr]{-7.70\\(-0.35\%)}
  & \makecell[tr]{-7.58\\(-0.30\%)} \\

Infer markings & 60
  & \makecell[tr]{-58.60\\(-3.30\%)}
  & \makecell[tr]{-101.82\\(-4.26\%)}
  & \makecell[tr]{-39.10\\(-1.76\%)}
  & \makecell[tr]{-76.52\\(-3.05\%)} \\

Semgrep markings & 60
  & \makecell[tr]{-61.00\\(-3.44\%)}
  & \makecell[tr]{-67.93\\(-2.84\%)}
  & \makecell[tr]{-43.70\\(-1.97\%)}
  & \makecell[tr]{-52.55\\(-2.10\%)} \\

Sink heuristic markings & 60
  & \makecell[tr]{-16.30\\(-0.92\%)}
  & \makecell[tr]{-16.68\\(-0.70\%)}
  & \makecell[tr]{-11.20\\(-0.50\%)}
  & \makecell[tr]{-16.58\\(-0.66\%)} \\

Sanitizer markings & 60
  & \makecell[tr]{-13.30\\(-0.75\%)}
  & \makecell[tr]{\textbf{+6.73}\\(\textbf{+0.28\%})}
  & \makecell[tr]{-3.40\\(-0.15\%)}
  & \makecell[tr]{-1.13\\(-0.05\%)} \\

No marking & 60
  & \makecell[tr]{-53.10\\(-2.99\%)}
  & \makecell[tr]{-63.08\\(-2.64\%)}
  & \makecell[tr]{-30.20\\(-1.36\%)}
  & \makecell[tr]{-50.43\\(-2.01\%)} \\

\midrule
\multicolumn{6}{@{}l}{\textbf{Prompt}} \\

Positive-only & 59
  & \makecell[tr]{-88.60\\(-4.99\%)}
  & \makecell[tr]{-138.25\\(-5.79\%)}
  & \makecell[tr]{-44.50\\(-2.00\%)}
  & \makecell[tr]{-102.52\\(-4.09\%)} \\

Negative-only & 59
  & \makecell[tr]{-63.00\\(-3.55\%)}
  & \makecell[tr]{-55.50\\(-2.32\%)}
  & \makecell[tr]{-42.10\\(-1.89\%)}
  & \makecell[tr]{-52.20\\(-2.08\%)} \\

\midrule
\multicolumn{6}{@{}l}{\textbf{Baseline}} \\

random-path\tnote{*} & 81
  & \makecell[tr]{-340.50\\(-19.18\%)}
  & \makecell[tr]{-301.14\\(-11.98\%)}
  & \makecell[tr]{-1158.20\\(-49.76\%)}
  & \makecell[tr]{-1087.52\\(-41.75\%)} \\

Empc & 70
  & \makecell[tr]{-567.40\\(-31.96\%)}
  & \makecell[tr]{-671.62\\(-28.12\%)}
  & \makecell[tr]{-1309.90\\(-58.91\%)}
  & \makecell[tr]{-1361.80\\(-54.29\%)} \\

CGS & 63
  & \makecell[tr]{-669.10\\(-37.69\%)}
  & \makecell[tr]{-321.90\\(-13.48\%)}
  & \makecell[tr]{-1285.40\\(-57.80\%)}
  & \makecell[tr]{-1095.43\\(-43.67\%)} \\

\bottomrule
\end{tabular}

\begin{tablenotes}
\footnotesize
\item Each coverage cell shows the absolute difference above its percentage difference.
\item[*] For \path{random-path}, both medians and averages are based on
10 benchmarks due to \textsc{KLEE} state exhaustion in \textit{jasper}
and \textit{readelf} before coverage collection; other configurations
use all 12 benchmarks. UV: unique security violation; BB: basic block.
\end{tablenotes}

\end{table}

\smallskip\noindent\textbf{Results.}
Table~\ref{tab:ablation-coverage-compare} presents four groups of comparisons, covering the searcher design, algorithmic components, marking sources, and prompt sensitivity, together with baseline strategies.

\textbf{Searcher Design.}
Compared with \tech{}, \textsc{SDSE} variant discovers substantially fewer unique violations, dropping from 87 to 54. All four coverage deltas are also negative, accounting for $-7.15\%$ to $-4.05\%$. Since \textsc{SDSE} variant uses the same LLM-generated markings as \tech{}, this result shows that shortest distance guidance cannot fully address the exploration bottleneck caused by cyclic control-flow regions, whereas \tech{}'s loop-exit path prioritization helps steer execution out of such regions and toward vulnerable paths that are difficult to reach.

\textbf{Internal Components.}
The two internal ablations show that both LLM-guided path prioritization and loop-exit path prioritization are necessary for \tech{}.

Without LLM-guided path prioritization, $v_1$ discovers only 61 unique violations, even though its coverage is higher than the full \tech{} across all four coverage metrics, from $+0.01\%$ to $+5.23\%$. This result shows that loop-exit path prioritization alone can improve general exploration by preventing \textsc{KLEE} from spending excessive budget in loop-heavy regions. However, the sharp decrease in unique violations (from 87 to 61) indicates that higher coverage does not necessarily translate into better vulnerability discovery. LLM-guided prioritization is needed to bias exploration toward security-relevant code.

Without loop-exit path prioritization, $v_2$ discovers only 58 unique violations, and all coverage deltas become negative, from $-6.01\%$ to $-3.60\%$. This result shows that LLM-generated markings alone are also insufficient. Even when target locations are semantically meaningful, \textsc{KLEE} can still spend its budget in cyclic regions and fail to reach the marked locations effectively. Overall, \tech{} needs both components: LLM-guided prioritization provides security-oriented targets, while loop-exit prioritization improves the ability to reach them.

\textbf{Marking Sources.}
The marking source comparisons show that replacing LLM-generated markings with random, static analysis, sink, sanitizer, or no markings consistently reduces unique security violation discovery, indicating that \tech{}'s effectiveness does not merely come from introducing additional target locations.

\begin{itemize}
\item \textbf{Random marking:}
Random markings discover only 61 unique violations, far below the 87 discovered by \tech{}. In terms of coverage, random markings have lower median deltas in both basic block coverage (-0.92\%) and line coverage (-0.51\%), but positive average deltas in basic-block coverage (+2.23\%) and line coverage (+1.15\%). This indicates that random markings are noisy: a few benchmark programs with large coverage gains raise the average, but the median remains lower than \tech{}. 

\item \textbf{Static analysis markings:}
These static analysis markings underperform \tech{}. CodeQL, Infer, and Semgrep each find only 60 unique violations, compared with 87 for \tech{}. Moreover, their coverage deltas are lower than \tech{} in both basic blocks and lines. This result is consistent with our motivation discussed in Section~\ref{sec:rw}: static analyzers rely on predefined rules, which may limit the potential vulnerability locations they can mark. Nevertheless, this result alone does not isolate why LLM-generated markings are more effective, which we further analyze in Section~\ref{sec:marking-quality}.

\item \textbf{Sink heuristic markings:}
The sink heuristic finds only 60 unique violations, compared with 87 for \tech{}. This result is expected because sink calls do not fully overlap with the security violations considered in this study. Blindly marking every call to a potentially risky sink function introduces substantial noise, because the safety of such calls is often determined by preceding checks rather than by the sink function itself. For example, in these experiments, the \texttt{write} function writes a specified number of bytes from a buffer to a file descriptor. A security violation may occur if the specified size exceeds the valid length of the buffer. However, in most cases, the size has already been safely computed, validated, or fixed as a constant by the preceding code. This suggests that markings derived only from predefined sink patterns are less effective than LLM-generated markings, which identify regions according to the prompt definition rather than fixed patterns.

\item \textbf{Sanitizer heuristic markings:}
\tech{} with sanitizer markings discovers only 60 unique violations. According to the markings, 60.05\% to 88.89\% of basic blocks are reachable across the benchmark programs. Under such dense markings, directed symbolic execution loses much of its discriminative guidance value. These markings guide directed symbolic execution almost randomly, resulting in a slight increase in average basic block coverage but making it difficult to spend the time budget approaching vulnerable code locations.

\item \textbf{No markings:}
The no-marking variant still discovers 60 unique violations despite achieving substantially lower coverage than the full \tech{}, indicating that loop-exit prioritization remains effective even without marking guidance.

\end{itemize}

\textbf{Prompt Sensitivity.}
Both positive-only and negative-only prompts discover only 59 unique violations, far below \tech{} with 87. Their coverage deltas are also negative across all four coverage metrics, from $-5.79\%$ to $-1.89\%$.

\begin{summarybox}
RQ4 shows that the \textsc{KLEE} searcher design, the searcher internal components, LLM-generated markings, and both positive and negative prompt examples make effective contributions to \tech{}. Compared with the 87 unique violations detected by the full version, the other ablation variants show a clear gap in security violation discovery, finding only 54 to 61 unique violations. These results show that \tech{} requires both effective path prioritization and high-quality markings to achieve the best result.
\end{summarybox}

\section{Discussion}
\subsection{Data Leakage Analysis}
\label{sec:llm_leak}
To mitigate potential data leakage in LLMs used for marking potentially vulnerable code regions, we assessed \tech{} in an open-source project created after the training data collection of \textit{DeepSeek-Coder-33B} (Feb. 2023~\citep{guo2024deepseekcoderlargelanguagemodel}). We searched candidates on GitHub using strict filters: \path{language:C}, \path{stars:>10}, and \path{archived:false}. For \textsc{KLEE} compatibility, we excluded projects involving multithreading, hardware, or non-Linux operating systems. The top result of this search was \path{zenc-lang/zenc}. The results show that \tech{} still outperforms most baselines in detecting security violations on this codebase, indicating \tech{} generalizes to new code and is not reliant on data seen during the LLM training phase. We also verified that 18.02\% of all markings generated by
\textit{DeepSeek-Coder-33B} across the 12 benchmarks correspond to source code lines introduced by commits after this training date.

\subsection{Code Coverage and LLM-Marked Regions Analysis}
To further examine how the coverage achieved by \tech{} relates to the regions marked by LLM, we conducted a case study on three representative benchmark programs, \texttt{jasper}, \texttt{tiffinfo}, and \texttt{transicc}. These were chosen because they show different coverage patterns compared to other benchmarks. In \texttt{jasper}, \tech{} achieves lower line and basic-block coverage. In \texttt{tiffinfo}, \tech{} improves line coverage but still shows lower basic-block coverage. In \texttt{transicc}, \tech{} exhibits inconsistent code coverage across five repeated runs.

Our manual inspection shows a consistent pattern across all three programs: the marked regions are located in the primary execution logic rather than in trivial functions. In \texttt{jasper}, the marked regions are mainly in deep JPEG2000 decoding and wavelet transform code. In \texttt{tiffinfo}, they cluster in directory traversal and tag-processing logic. In \texttt{transicc}, they focus on internal type handling and interpolation code. This shows that \tech{} does identify code that is critical to the program's main behavior.

This case study also shows that marking important code does not always lead to the same coverage gain. The marked code in \texttt{jasper} is embedded in deeply nested mathematical logic, \texttt{tiffinfo} involves broad metadata traversal, and \texttt{transicc} centers on more localized data transformation pipelines. In those cases, the marked code is simply harder to reach and explore due to its structure.

\subsection{Token and Time Cost Analysis}

\begin{table*}[!t]
\centering
\caption{Token usage of LLM-generated markings.}
\label{tab:llm-token-usage}

\resizebox{\columnwidth}{!}{
\begin{tabular}{@{}p{94pt}rrrrrr@{}}
\toprule
\textbf{Model} &
\makecell[r]{\textbf{Input}\\\textbf{Tokens}} &
\makecell[r]{\textbf{Output}\\\textbf{Tokens}} &
\makecell[r]{\textbf{Average /}\\\textbf{Program}} &
\makecell[r]{\textbf{Median /}\\\textbf{Program}} &
\makecell[r]{\textbf{Token per}\\\textbf{KLOC}} &
\makecell[r]{\textbf{Total}\\\textbf{Price}} \\
\midrule

DeepSeek-Coder-6.7B
& 21,352,114 & 1,046,322
& 1,866,536 & 736,244 & 29,756 & -- \\

DeepSeek-Coder-33B
& 21,212,891 & 883,713
& 1,841,384 & 735,183 & 29,355 & -- \\

Qwen2.5-Coder-7B
& 20,600,403 & 411,091
& 1,750,958 & 677,950 & 27,913 & -- \\

Qwen2.5-Coder-32B
& 20,197,602 & 742,355
& 1,744,996 & 557,640 & 27,818 & -- \\

Gemini-2.5-Flash
& 13,408,173 & 895,639
& 1,191,984 & 968,364 & 19,002 & \$6.26 \\

DeepSeek-V4-Flash
& 9,552,223 & 1,270,727
& 901,912 & 714,312 & 14,378 & \$1.69 \\

GPT-4o
& 8,124,113 & 409,613
& 711,144 & 502,274 & 11,337 & \$24.41 \\

\bottomrule
\end{tabular}
}
\begin{tablenotes}[flushleft]
\footnotesize
\item Average and median are computed over the 12 evaluated programs.
Token per KLOC is computed using the code-line counts; the total
benchmark size is 752,749 LOC (752.749 KLOC).
Total price is computed only for cloud models using API list prices
and is reported in U.S. dollars (USD).
\end{tablenotes}

\end{table*}

\textbf{Token Cost.} A concern is the monetary cost of LLM-generated markings. In our study, we primarily use open-source models deployed locally, and also use three cloud models through APIs. Table~\ref{tab:llm-token-usage} reports the measured token usage and monetary cost for each LLM. Most tokens come from source code and our defined prompt as inputs, while the generated markings are relatively small. Therefore, for cloud models, the monetary cost is mainly determined by input token usage.
Across all models, average token usage per program is higher than the median. Compared with the local models, the cloud models use fewer tokens on average per program and fewer tokens per KLOC, with GPT-4o having the lowest token usage on both measures.

We also compute the cloud model cost and all monetary costs reported in U.S. dollars (USD). Under this accounting, marking all 12 benchmark programs costs USD~1.69 with \textit{DeepSeek-V4-Flash}, USD~6.26 with \textit{Gemini-2.5-Flash}, and USD~24.41 with \textit{GPT-4o}. These costs are small relative to the benefit of the resulting guidance. \textit{GPT-4o} and \textit{DeepSeek-V4-Flash} each find 87 unique security violations, matching the best local model markings, \textit{DeepSeek-Coder-33B}. Therefore, the best result does not require an expensive cloud model.

\textbf{Time Cost.} We further measure the time spent on generating markings for the 14,626 functions in the 12 benchmark programs. As shown in Table~\ref{tab:llm-time-cost}, the cloud models are substantially faster, taking 0.28--0.99 seconds per function, whereas the local models take 0.75--3.46 seconds per function. Note that all local models are deployed on a single consumer-grade NVIDIA RTX 3090 GPU, and newer GPUs are expected to further reduce their marking time. Moreover, \textit{DeepSeek-Coder-6.7B} discovers the same 87 unique violations as \textit{DeepSeek-Coder-33B} (Section~\ref{sec:rq3_results}) while taking only 1.48 seconds per function, less than half of the time of \textit{DeepSeek-Coder-33B}. This suggests that a small local model is sufficient for \tech{} on this task.

\begin{table*}[t]
\centering
\caption{Time cost of LLM-generated markings per function.}
\label{tab:llm-time-cost}

\begin{tabular}{@{}lr@{}}
\toprule
\textbf{Model} & \textbf{Time per Function (seconds)} \\
\midrule
DeepSeek-Coder-6.7B & 1.48 \\
DeepSeek-Coder-33B  & 3.46 \\
Qwen2.5-Coder-7B    & 0.75 \\
Qwen2.5-Coder-32B   & 3.27 \\
DeepSeek-V4-Flash   & 0.99 \\
Gemini-2.5-Flash    & 0.32 \\
GPT-4o              & 0.28 \\
\bottomrule
\end{tabular}
\end{table*}

\subsection{LLM-Generated Marking Quality Analysis}
\label{sec:marking-quality}
Section~\ref{sec:rq4_results} shows that LLM-generated markings improve violation discovery, but not how these markings relate to the discovered violations. Because the root cause of a violation and the location where a sanitizer reports it often differ~\citep{blazytko2020aurora}, we do not require a marking to overlap with the reported location. Instead, we consider the top $N$ stack frames of each violation, which are associated with triggering the same bug, with $N$ often between 3 and 5~\citep{klees2018evaluating}. Each stack frame represents a function invocation, and its position in the call stack indicates its call distance from the reported violation. A violation is associated with a marking source if any of these frames contains a marking from that source. Table~\ref{tab:marking-association} shows that the markings generated by \textit{DeepSeek-Coder-33B} are associated with 69\% and 71\% of the unique violations discovered by \tech{} for $N=3$ and $N=5$, respectively, substantially higher than CodeQL, Infer, and random markings.

\begin{table*}[t]
\centering
\caption{Proportion of unique violations associated with markings in the top $N$ stack frames.}
\label{tab:marking-association}

\begin{tabular}{@{}lrrrr@{}}
\toprule
\textbf{Top Frames} &
\textbf{DeepSeek-Coder-33B} &
\textbf{CodeQL} &
\textbf{Infer} &
\textbf{Random} \\
\midrule
$N = 3$ & \textbf{69\%} & 21\% & 11\% & 26\% \\
$N = 5$ & \textbf{71\%} & 51\% & 28\% & 44\% \\
\bottomrule
\end{tabular}

\end{table*}

We further manually assess the precision of LLM-generated markings by randomly sampling 92 of the 2,031 function-level markings generated by \textit{DeepSeek-Coder-33B}, which ensures a 95\% confidence level with a 10\% margin of error. Consistent with the definition of security violations in our prompt (Section~\ref{sec:phase1}), a marking is regarded as relevant if the marked code may lead to integer-related undefined behavior, integer-driven boundary or length violations, or memory safety violations such as use-after-free and buffer overflows. Overall, 59 of the 92 markings (64.1\%) are relevant. Among the 33 irrelevant markings, 10 cover only non-executable lines (e.g., function signatures and braces), and the remaining 23 cover logging, accessors, control-flow statements, or constant data. Since markings only guide path prioritization rather than report vulnerabilities, such false positives do not produce false alarms but only redirect part of the exploration budget.

\subsection{Missed Violations Analysis}
\label{sec:missed-violations}

\begin{table*}[t]
\centering
\caption{Violations exclusively discovered by baselines and missed by \tech{}.}
\label{tab:missed-violations}

\begin{tabular}{@{}llrrr@{}}
\toprule
\textbf{Program} &
\textbf{Strategy} &
\makecell[r]{\textbf{\# Exclusive}\\\textbf{Violations}} &
\makecell[r]{\textbf{Loop}\\\textbf{Iterations}} &
\makecell[r]{\textbf{Found}\\\textbf{by $v_2$}} \\
\midrule
\texttt{nasm}     & \path{dfs}     & 4 & 1 & 0 \\
\texttt{nasm}     & \textsc{Empc}  & 2 & 6 & 0 \\
\texttt{transicc} & \path{dfs}     & 2 & 1 & 0 \\
\texttt{flvmeta}  & \path{nurs:rp} & 2 & 1 & 0 \\
\bottomrule
\end{tabular}
\end{table*}

As shown in Figure~\ref{fig:crash-overlap}, \path{dfs}, \textsc{Empc}, and \path{nurs:rp} discover 6, 2, and 2 exclusive violations missed by \tech{}, respectively. Because loop-exit prioritization deprioritizes cyclic regions without LLM-generated markings, a missing marking may be amplified into a missed violation. To examine this interaction, we replay the test cases of these 10 violations to count the loop iterations required to trigger them, and check whether they are discovered by $v_2$ (without loop-exit prioritization) in Section~\ref{sec:rq4_results}. Table~\ref{tab:missed-violations} summarizes the results.

These violations are located in features such as response-file processing in \texttt{nasm} and XML dumping in \texttt{flvmeta}, and 8 of them are triggered in the first loop iteration. Although $v_2$ disables loop-exit prioritization, it still cannot discover any of them. The root cause is that neither these violation locations nor the nodes dominated by them in the ICFG contain LLM-generated markings, so entering these regions does not shorten the distance to any marking, and they are therefore not explored. This reflects an inherent trade-off of directed exploration, where regions without targets receive lower priority, rather than an amplification by loop-exit prioritization.

\begin{figure}[!htbp]
\centering
\begin{lstlisting}[style=cviol]
static const char *debug_format;
/* process_arg(): option -g read from a response file */
debug_format = nasm_skip_spaces(p + 2);   /* points into the line buffer */
/* process_respfile(): end of the response file */
(*@\viol{nasm\_free(buffer);}@*)
/* main(): debug format lookup */
dfmt = dfmt_find(ofmt, (*@\viol{debug\_format}@*));     /* heap-use-after-free */
\end{lstlisting}
\caption{A heap-use-after-free in \texttt{nasm} exclusively discovered by dfs}
\label{fig:nasm-uaf}
\end{figure}

Figure~\ref{fig:nasm-uaf} illustrates the failure mode with a representative case. When the option \texttt{-g} is read from a response file, \texttt{nasm} stores a pointer into the line buffer in \texttt{debug\_format}, frees the buffer at the end of the file, and later dereferences the dangling pointer. Triggering this violation requires a recursive call: \texttt{process\_arg} handles the option \texttt{-@} by calling \texttt{process\_respfile}, which in turn calls \texttt{process\_arg} for each line of the response file. A recursive call is a special form of cyclic structure in a program: in the ICFG, the call and return edges among the recursive functions form a strongly connected component, just as a loop does. Neither this cyclic region nor the nodes dominated by it contain LLM-generated markings. Therefore, \tech{} has no target to steer into the recursion, even when loop-exit prioritization is disabled in $v_2$. In contrast, \path{dfs} continuously follows the most recently forked states and thus naturally descends into the recursive call.

\subsection{Fallback Threshold Analysis}
\tech{} returns to random selection after executing 35 million instructions without covering any new basic block (C4). To examine whether our results depend on this threshold, we measure, for each run of \tech{}, the longest interval (in instructions) without covering any new basic block. Table~\ref{tab:fallback-stall} shows that the longest interval across all 60 runs is 16.47 million instructions, observed on \texttt{transicc}, and thus the fallback is never triggered. This suggests that the fallback acts only as a safeguard rather than a tuned component, and that the effectiveness of \tech{} is insensitive to the choice of this threshold.

\begin{table*}[t]
\centering
\caption{Coverage stalls across five runs (million instructions).}
\label{tab:fallback-stall}

\begin{tabular}{@{}lrrr@{}}
\toprule
\textbf{Program} &
\textbf{Average} &
\textbf{Median} &
\textbf{Max} \\
\midrule
bc        & 0.08  & 0.08  & 0.10  \\
tic       & 4.79  & 4.53  & 5.69  \\
bison     & 0.08  & 0.08  & 0.10  \\
readelf   & 2.95  & 2.97  & 3.00  \\
strip-new & 4.94  & 4.99  & 5.32  \\
make      & 0.00  & 0.00  & 0.00  \\
nasm      & 4.23  & 4.13  & 5.36  \\
tiffinfo  & 2.92  & 2.73  & 3.33  \\
jasper    & 1.47  & 1.25  & 2.60  \\
transicc  & 14.80 & 15.05 & 16.47 \\
flvmeta   & 0.50  & 0.48  & 0.59  \\
curl      & 9.63  & 9.57  & 9.83  \\
\midrule
\textbf{All} &
\textbf{3.87} &
\textbf{2.96} &
\textbf{16.47} \\
\bottomrule
\end{tabular}
\end{table*}

\section{Threat to Validity}

\textbf{Internal Validity.}
First, because \textsc{CGS} and \textsc{Learch} depend on original versions of \textsc{KLEE} and LLVM, this setup may affect our results. Since updating these baselines could alter their initial behavior~\citep{empc2025}, we chose to adopt \textsc{Empc}'s evaluation setup. 
Second, the code coverage in \textsc{KLEE} may vary across runs. To mitigate this threat, following previous work~\citep{he2021learch,empc2025}, we repeat each experiment five times and report the mean with one standard deviation. In our results, noticeable variance appears on only a small number of benchmarks.
Third, LLMs exhibit non-determinism, which can affect the identification of vulnerable locations and, consequently, the performance of \tech{}. To mitigate this threat, we set the temperature to 0 and use two general-purpose models with non-thinking settings. We additionally run \textit{DeepSeek-Coder-33B}, our primary model, five times with caching disabled, and obtain identical markings across all runs.
Fourth, loop-exit prioritization may still introduce false negative risk when the LLM fails to mark a vulnerable location inside a cyclic region, which we acknowledge as a trade-off in our design.

\textbf{Construct Validity.}
First, we may overestimate our approach, as we use the violation-discovery metric and overlook certain aspects (e.g., low-level errors and user-defined assertion violations~\citep{cadar2008klee}). To ensure a fair comparison, we follow recent studies~\citep{empc2025, sun2024cgs, he2021learch} that use the detection of security violations as a key evaluation metric.
Second, validation of security violations by replaying could be unstable as sanitizer-instrumented executions may behave differently from regular executions~\citep{asan-wiki}. Therefore, in each experiment, we replay each test case five times to improve stability, with a 10-minute timeout for each replay to ensure completeness, consistent with previous studies~\citep{empc2025,he2021learch}.
Third, we cannot compute classical precision and recall for LLM-generated markings because the benchmark programs adopted from prior studies~\citep{empc2025, sun2024cgs, he2021learch} do not provide a complete ground truth of all vulnerable locations. A marking that does not lead to a confirmed security violation within the symbolic execution budget may simply remain unexplored, rather than being a false positive. We therefore conduct a complete ablation study to evaluate the performance improvement brought by LLM-generated markings.
Fourth, potential LLM data leakage remains a concern as LLMs are heavily trained on open-source repositories; LLMs could already have memorized those security violations in our benchmark set. To mitigate this, we conducted additional experiments in Section \ref{sec:llm_leak}, confirming the effectiveness of \tech{}.

\textbf{External Validity.} 
First, our evaluation is conducted only on a set of 12 benchmarks, which are widely-used real-world applications spanning six distinct functional types from a previous study~\citep{empc2025}, to ensure fair comparisons with other baselines. Therefore, our results may not generalize to other programs.  Second, the detection capability of \tech{} is constrained by ASan and UBSan, which serve as the primary security violation verification mechanisms in \textsc{KLEE}. In future work, we plan to extend our approach beyond the \textsc{KLEE} to cover more diverse security violation types. Third, the choice of models employed for vulnerable markings could affect the effectiveness of \tech{}. We mitigate this by conducting a model sensitivity analysis including four open-source coder models, three general-purpose models widely used in recent security studies~\citep{lin2025comparative,ding2025vulnerability,sun2024llm4vuln,bae2024gpt4o_vuln,bruni2025forge_gpt4o,siddiq2025security_comprehension,qin2025reasoning_vuln_eval}. Due to the resource limitations, it is not feasible to evaluate all available models.

\section{Conclusion and Future Work}
This paper presents \tech{}, which integrates the \textsc{KLEE} symbolic execution engine with LLM-marked locations for security violation discovery. \tech{} addresses the path explosion problem, the primary bottleneck of symbolic execution from two perspectives: (i) LLM-guided path prioritization, which directs execution toward potential vulnerabilities, and (ii) a loop-exit path prioritization that avoids being trapped in loops. By incorporating code-semantic context into path prioritization, our approach outperforms existing strategies in both code coverage and security violation discovery.

Despite these results, two constraints remain for future research. First, the types of detectable violations and supported programming languages are limited by the underlying capabilities of \textsc{KLEE} and compiler sanitizers. A key future direction is to develop mechanisms that can be applied across more diverse software ecosystems. Second, as a multi-path symbolic execution engine, \textsc{KLEE} cannot always concentrate its budget on a specific path. Factors such as SMT solver timeouts and incomplete external symbols may cause promising paths to be incorrectly treated as infeasible. In future work, we aim to bypass these architectural constraints by shifting symbolic execution to be more directed at security violations.

\section*{Statements and Declarations}

\textbf{Funding}\\
We gratefully acknowledge the financial support of: (1) JSPS for the KAKENHI grants (JP25K22845, JP26H02500, and JP26K21198); (2) Japan Science and Technology Agency (JST) as part of Adopting Sustainable Partnerships for Innovative Research Ecosystem (ASPIRE), Grant Number JPMJAP2415, (3) the Kayamori Foundation of Informational Science Advancement for supporting Tao Xiao, and (4) the Inamori Research Institute for Science for supporting Yasutaka Kamei via the InaRIS Fellowship.\\

\noindent
\textbf{Ethical Approval}\\
Not applicable.\\

\noindent
\textbf{Informed Consent}\\
Not applicable.\\

\noindent
\textbf{Author Contributions}\\
Conceptualization: Lingfeng Chen and Tao Xiao; Methodology: Lingfeng Chen and Tao Xiao; Formal analysis and investigation: Lingfeng Chen and Tao Xiao; Writing original draft preparation: Lingfeng Chen; Writing review \& editing: Lingfeng Chen, Tao Xiao, Masanari Kondo, and Yasutaka Kamei; Resources: Tao Xiao and Yasutaka Kamei; Supervision: Tao Xiao, Masanari Kondo, and Yasutaka Kamei. All authors read and approved the final manuscript. \\

\noindent
\textbf{Data Availability Statement}\\
Our replication package is available online~\citep{repulication_package} and includes (i) the source code of \tech{} and baselines; (ii) experimental results; (iii)  reported violations; and (iv) a Dockerized benchmark suite.\\

\noindent
\textbf{Competing Interests}\\
The authors declare that Yasutaka Kamei is a member of the EMSE Editorial Board. All co-authors have seen and agree with the contents of the manuscript, and there are no financial interests to report.\\

\bibliography{sample-base}


\end{document}